\documentclass[prl,twocolumn,showpacs,amsmath,amssymb,superscriptaddress]{revtex4-2}
\usepackage{graphicx}
\usepackage{hyperref}
\usepackage{xcolor}
\usepackage{color}
\usepackage[all,cmtip]{xy}
\usepackage{amsmath,amsthm} 
\usepackage{amsfonts}

\makeatother

\begin{document}

\title{Topological electronic structure and Weyl points in nonsymmorphic hexagonal materials}
\author{Rafael Gonz\'{a}lez-Hern\'{a}ndez}
\email{rhernandezj@uninorte.edu.co}
\affiliation{Departamento de F\'{i}sica y Geociencias, Universidad del Norte, Km. 5 V\'{i}a Antigua Puerto Colombia, Barranquilla 080020, Colombia}
\affiliation{Institut f\"ur Physik, Johannes Gutenberg Universit\"at Mainz, D-55099 Mainz, Germany}
\author{Erick Tuiran}
\email{etuiran@uninorte.edu.co}
\affiliation{Departamento de F\'{i}sica y Geociencias, Universidad del Norte, Km. 5 V\'{i}a Antigua Puerto Colombia, Barranquilla 080020, Colombia}
\author{Bernardo Uribe}
\email{bjongbloed@uninorte.edu.co}
\affiliation{Departamento de Matem\'{a}ticas y Estad\'{i}stica, Universidad del Norte, Km. 5 V\'{i}a Antigua Puerto Colombia, Barranquilla 080020, Colombia}

\date{\today}

\begin{abstract}

Using topological band theory analysis we show that the nonsymmorphic symmetry operations in hexagonal lattices enforce Weyl points at the screw-invariant high-symmetry lines of the band structure. The corepresentation theory and connectivity group theory show that Weyl points are generated by band crossings in accordion-like and hourglass-like dispersion relations. These Weyl points are stable against weak perturbations and are protected by the screw rotation symmetry. Based on first-principles calculations we found a complete agreement between the topological predicted energy dispersion relations and real hexagonal materials. Topological charge (chirality) and Berry curvature calculations show the simultaneous formation of Weyl points and nodal-lines in 4$d$ transition-metal trifluorides such as AgF$_3$ and AuF$_3$. Furthermore, a large intrinsic spin-Hall conductivity was found due to the combined strong spin-orbit coupling and multiple Weyl-point crossings in the electronic structure. These materials could be used to the spin/charge conversion in more energy-efficient spintronic devices.

\end{abstract}
\maketitle

\section{Introduction}

The intersection between band-theory of solids and topology has recently attracted a lot of interest in the condensed matter physics community. In the past decade, topological insulators opened the door to novel phases of matter with unique properties \cite{Colloquium-topological-insulators}. The field of topological materials has rapidly expanded and new kind of topological phases have appeared, including Chern insulators \cite{Haldane-model}, topological insulators \cite{Kane-Mele-model}, crystalline topological insulators \cite{Topological-Crystalline-Insulators}, Dirac semimetals \cite{Discovery-Dirac-Semimetal}, Weyl semimetals \cite{Weyl-and-dirac-semimetals} and nodal-line semimetals \cite{Topological-nodal} among others. Weyl semimetals have been the subject of intensive research because they were the first topological material found without a bulk energy gap which is protected by the non-trivial topology of the band structure \cite{Weyl-and-dirac-semimetals}. The existence of Weyl semimetals was predicted theoretically about ten years ago and was finally found experimentally in 2015 \cite{Discovery-of-Weyl-semimetals,Discovery-Weyl-fermion}.

It is theoretically known that a Weyl semimetal can only arise in a crystal where time-reversal symmetry or inversion symmetry are broken \cite{Symmetry-demanded-topological}. For these cases the Hamiltonian around the Weyl point could be described in such a way that the energy crossings are linearly dispersing energy-bands which act as monopoles of Berry curvature in momentum space \cite{Weyl-and-dirac-semimetals}.  These monopoles are characterised by a quantised topological charge (or chirality) and the total chiral charge must be zero (this result is known as the Nielsen-Ninomiya theorem \cite{Absence-of-neutrinos-in-lattice}). However, the space group symmetries in solids can be complicated and high-order Weyl crossings \cite{Multi-Weyl-Topological-Semimetals,Symmetry-Protected-Topological-Triangular-Weyl-Complex} or Weyl points and nodal-lines combinations can be present at the material.

The variety of topological band crossing in materials is the essential ingredient that guarantees the existence of the novel charge and spin transport properties that Weyl semimetals show \cite{Observation-Chiral-Anomaly-Induced}.  Therefore, the generation of new materials with stable Weyl points is an emergent field of great interest from a fundamental point of view and possible technological applications.  Weyl points in the band structure can be emergent by the nonsymmorphic symmetries of the crystal; these Weyl points are commonly called symmetry-enforced energy band crossings. Previous works have investigated several nonsymmorphic energy band crossings \cite{Nonsymmorphic-symmetry-required-band-crossings,semimetals-in-nonsymmorphic-lattices, PhysRevB.96.045102} and the possible materials realisation \cite{Tunable-Weyl-and-Dirac-states,Weyl-semimetal-pyrochlore-oxides}. In particular, Weyl points protected by screw rotations and Weyl nodal lines protected by glide reflections were reported recently in trigonal and hexagonal lattices \cite{Topological-crossings-trigonal,Topological-crossings-hexagonal}.

In this work, we study the generation of Weyl points due to the screw nonsymmorphic symmetry in hexagonal systems with the P$6_p$ and P$6_p$22 space groups. First, based on corepresentation theory \cite{Wigner} and band connectivity group theory \cite{The-mathematical-theory-of-symmetry-in-solids}, we study the combinatorics of the complete electronic band structure. We note the appearance of accordion-like and hourglass-like energy dispersion relations inducing the existence of Weyl points protected by the screw and time-reversal symmetries. This analysis appears in Appendix A.  Second, using first-principles calculations we corroborate the energy band crossings for hexagonal materials. In particular we analyse the electronic band structure of the materials In$_2$Se$_3$, KCaNd(PO$_4$)$_2$, PI$_3$, AgF$_3$, AuF$_3$, TaGe$_2$ and Nb$_3$CoS$_6$ with the refined structural parameters from the Materials project and AFLOW database \cite{materialsproject,aflow}. This appears in Appendix B. The content of both appendices A and B expands and enhances the previous analysis carried out in
\cite{Topological-crossings-hexagonal}. In particular we have added the analysis of the energy bands along the high symmetry line K-H and we have described a complete combinatorial band 
structure that appears on materials with symmetry groups P6$_3$ and P6$_1$22.
 Third, we use the topological band analysis to deduce and infer topological properties of the Weyl points in AgF$_3$ and AuF$_3$ materials. These materials are possible Weyl-semimetal candidates with inversion symmetry breaking. Following the Nielsen-Ninomiya theorem, we predict the presence of both Weyl points and nodal lines at AgF$_3$ and AuF$_3$. We find interesting distributions of Weyl points as it is the case of the second and third conduction bands of AgF$_3$ where we find 6 Weyl points with $+2$ chirality, 24 points with $+1$ chirality and $36$ points with $-1$ chirality. We also show the existence of Weyl nodal lines on the tenth and eleventh conduction bands of AgF$_3$. This analysis of the distribution of Weyl points allows us to confirm that the Weyl points do not necessarily come in pairs with opposite chirality \cite{Symmetry-Protected-Topological-Triangular-Weyl-Complex}, and furthermore, that Weyl nodal lines exist whenever the distribution of  Weyl points along the high symmetry lines and their total chirality satisfy a numerical condition. Finally, we analyse the spin transport properties and we find a strong spin-Hall effect due to the presence of Weyl fermions in the valence band of AgF$_3$ and AuF$_3$ materials. This interesting spin-transport property in 4d transition-metal fluorides may help in the quest for the use of Weyl nodes in developing the next-generation of energy-efficient information technology  \cite{Composite-topological-nodal-lines-AgF2}.

\section{Electronic structure of  A\lowercase{g}F$_3$ and A\lowercase{u}F$_3$ (P6$_1$22)}

Let us start by focusing on the materials that we have studied. We have
chosen the materials AgF$_3$ and AuF$_3$ because they are both possible
Weyl-semimetals with the same nonsymmorphic hexagonal symmetry (P6$_1$22)
generated by a sixfold screw rotation along the $z$-axis for a hexagonal
lattice and a $2$-fold rotation. Due to the lack of inversion, mirror, and
roto-inversion symmetries these materials can be classified as chiral
crystals \cite{B}. In particular, we have chosen the right-handed P6$_1$22
space group, which is an enantiomorphic variant of P6$_5$22 space group
(left-handed). In appendix A, we have shown that these enantiomorphic pairs
(P6$_1$22 and P6$_5$22) present energetically degenerate band structures,
which is in agreement with the work by Li \emph{et al} \cite{Chiral-fermion-reversal}.
However, it has been recently shown that the topological charges and surface
Fermi arc can be reversed for the crystal with opposite enantiomer, which
can be measured by chiral optical response experiments \cite{Quantized-circular,Chiral-optical-response}.

The electronic band structures of AgF$_3$ and AuF$_3$ materials have similar
features and the topological properties of their conduction and valence
bands could be understood via topological band analysis of the symmetry
group P6$_1$22. If one takes a close look at the valence and conduction band
structures that could be seen in Figures \ref{AgF3-berry}a), \ref{AuF3-berry}a) and \ref{P6122-bandas} one notices that there are interesting crossings at the high symmetry lines $\Gamma$-A, K-H and M-L. These crossings appear due to the nonsymmorphicity of the screw rotation and could be understood
via a topological band analysis of the corepresentations of the high
symmetry points and of the high symmetry lines. We have carried out a
comprehensive study of these topological band structures for the symmetry
groups P6$_p$ and p6$_p$22 in Appendices A and B, and we have shown that
these hour-glass and accordion-like band structures on the high symmetry
lines are unavoidable and therefore they are topologically protected by the
symmetry.
 
Our interest is focused on the band crossings that appear in the interior of the high symmetry lines and not on the crossings that do exist on the high symmetry points which produce Kramers-Weyl points as has been done in references \cite{B,C,D, Sanchez2019, Rao2019, PhysRevLett.122.076402, Schroter2019}. Here we have focused on studying a pair of bands that do cross inside the
high-symmetry lines and the electronic properties that they induce on the hexagonal materials \cite{Topological-crossings-hexagonal}. These crossings are globally stable and can not be taken off by large perturbation of the system. In particular, we will concentrate our study on enforced band crossings of the third and second to last valence bands on AgF$_3$ and AuF$_3$ and on five-pairs of conduction bands of AgF$_3$.

\subsection{Weyl points on valence bands of AgF$_3$}
\begin{figure*}
	\includegraphics[width=17cm]{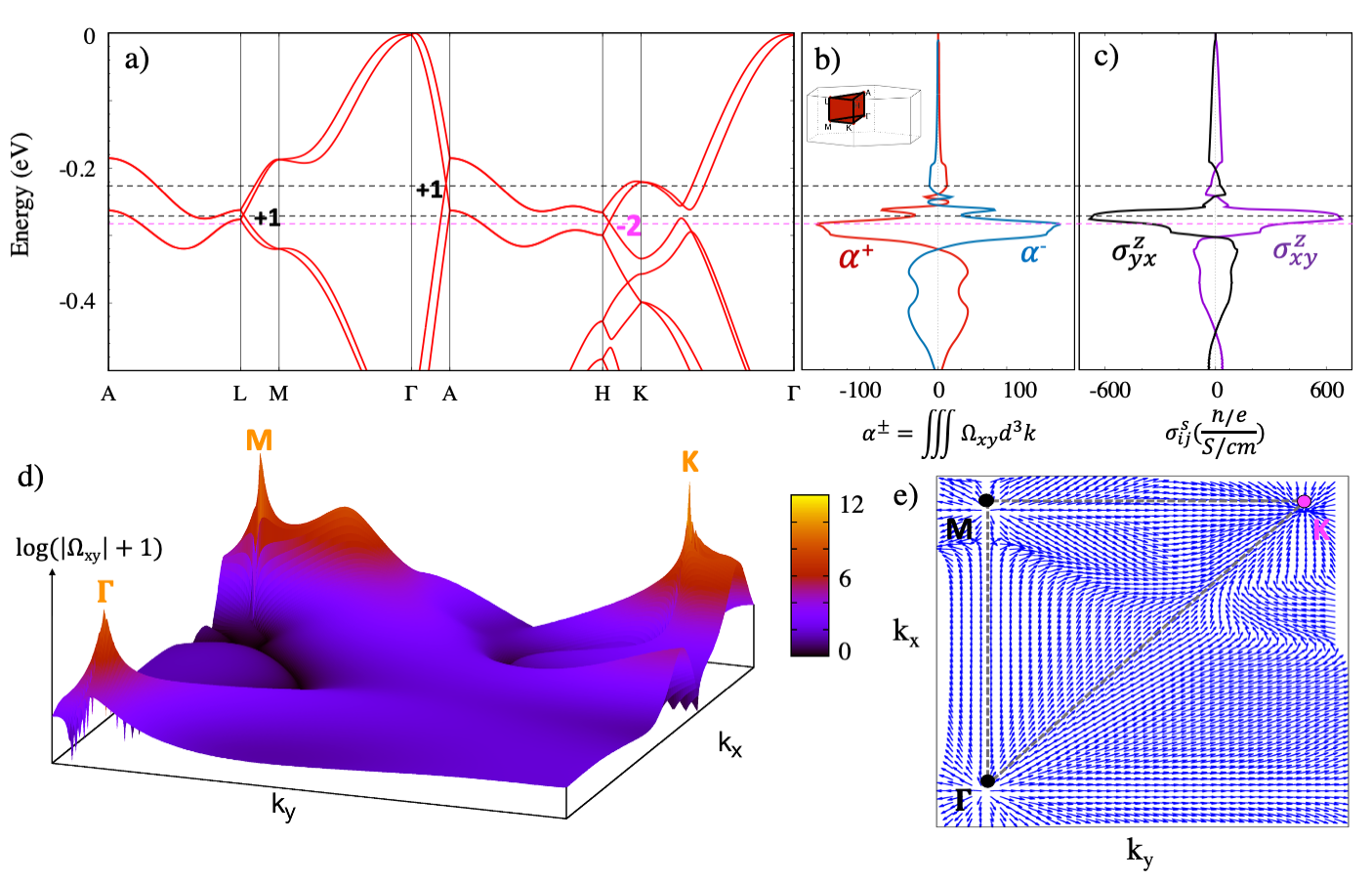}
	\caption{a) Valence band structure and chirality for the three Weyl points shown between the high-symmetry $k$-lines (L-M, $\Gamma$-A, H-K) of the BZ in AgF$_3$. b) Integration of the $xy$-component of Berry curvature ($\Omega_{xy}$) on the $k_z$-positive ($\alpha^{+}$) and $k_z$-negative ($\alpha^{-}$) regions of the irreducible first BZ as a function of energy. c) $\sigma_{xy}^{z}$ and $\sigma_{yx}^{z}$ components of the spin Hall conductivity tensor as a function of energy. d) $\log(|\Omega_{xy}|+1)$  on the $k$-plane touching the three Weyl points.   e) Top view of the Berry curvature field on the $k$-plane touching the three Weyl points. M and $\Gamma$ are sources ($+1$ chirality) while K is a sink ($-2$ chirality). Here M, $\Gamma$ and K represent the location of Weyl points at the $k_z$-lines L-M, $\Gamma$-A and H-K respectively. The Fermi level is set to zero.} \label{AgF3-berry}
\end{figure*}

The electronic band structure for this material below the Fermi level is shown in Figure \ref{AgF3-berry}a), where it is shown also the value of the chirality ($+1$ and $-2$) for the Weyl points close to the maximum of the valence band energy and that belong to the third and second to last bands. From Figure  \ref{AgF3-berry}a) we note that close to the valence band maximum (0 eV) it is possible to find three different Weyl points with close energy values of -0.220, -0.278 and -0.267 eV (dotted lines in Figure \ref{AgF3-berry}(a)). These nodes live along the $k_z$-paths $\Gamma$-A, K-H, and M-L and they have also similar $k_z$ components 0.449, 0.414, and 0.463 respectively (in fractional coordinates of the $k_z$-line). In particular, we have calculated the position of these nodal points using a gradient conjugate technique for a $k$-mesh grid of the Brillouin zone (BZ)  \cite{wanniertools} and these locations are consistent with the group theoretical predictions (see Diagram  \ref{complete topological band structure P6122}). 

 Weyl chirality calculations (see Appendix A) reveals that there are two Weyl points of chirality $+1$ induced by the Weyl point along the $\Gamma$-A path, 6 points of chirality $+1$ induced by the Weyl point along the M-L path and 4 points of chirality $-2$ induced by the Weyl point along the K-H path (See Figure  \ref{AgF3-berry}a)).  The sum $2+6-8$ of those chiralities is zero and Nielsen-Ninomiya theorem \cite{Absence-of-neutrinos-in-lattice} is satisfied. We have also checked that there are no other Weyl points present besides the ones previously presented.



\subsection{Transport properties of AgF$_3$}

In order to check the local manifestation of the geometric properties of the wave-functions in $k$-space, we have calculated the $xy$-component of the Berry curvature ($\Omega_{xy}$) on the $k$-plane that defines the above mentioned three Weyl points in $\Gamma$-A L-M and H-K respectively, as it is shown in Figure \ref{AgF3-berry}d). We can see a large contribution of the Berry curvature near to degeneracy points, where M, $\Gamma$ and K represent the location of Weyl points at the $k_z$-lines L-M, $\Gamma$-A and H-K respectively. In addition, the Berry curvature field indicates that among these Weyl points $\Gamma$ and M act as `sources' and K as `sink' of the Berry curvature field in the momentum space (see Figure \ref{AgF3-berry}e)). These topological monopole charges are located at the (degenerate) band crossings and their signs are consistent with the chirality calculated $\chi$=+1 for $\Gamma$-A and M-L and $\chi$=-2 for K-H (see Figure \ref{AgF3-berry}a)). These Weyl points are protected by the nonsymmorphic crystal symmetry and time-reversal symmetry presented in the AgF$_3$ (P6$_1$22) material.  These points are near to the valence band maximum (around 250 $m$eV) and they could be reached with a strong $p$-doping of this semiconductor material. 

On the other hand, the integration of the $xy$-component of Berry curvature ($\Omega_{xy}$) on the $k_z$-positive ($\alpha^{+}$) and $k_z$-negative ($\alpha^{-}$) regions of the first BZ as a function of energy is shown in Figure \ref{AgF3-berry}b). We found a high partial contribution to the anomalous Hall conductance (AHC) due to the K-H Weyl points (around -0.278 eV) even though the total AHC is zero due to time-reversal symmetry (the positive $\alpha^{+}$ and negative $\alpha^{-}$ integration on the full BZ cancel). This phenomenon could be understood due to different chirality values of the Weyl points close to the M-L and K-H lines, thus increasing and decreasing the AHC in a small range of energy.  It is known that positive and negative peaks on the electrical Hall conductivities can be sources of considerable large spin Hall conductivity (SHC) \cite{Bernevig2006}. We found that the total integral of the spin Hall curvature, which is even in TRS, is not zero and its huge contribution can be attributed to the presence of Weyl points in the valence band. From Figure \ref{AgF3-berry}c) we can see a large value for the spin Hall conductivity (SHC) of around 600 ($\hbar$/e)(S/cm), far larger than the one of the pure element Ag which lies around 100 ($\hbar$/e)(S/cm)  \cite{SpinHallEffects-Sinova}. These results are in complete agreement with the crystal symmetry analysis for the full SHC tensor for the $P6_122$ space group, on which the $\sigma_{xy}^{z}$ ($-\sigma_{yx}^{z}$) is not zero \cite{Seemann2015}.  Therefore we propose that AgF$_3$ can exhibit a large intrinsic spin Hall effect (SHE) mainly due to the particular contribution of the Weyl points to the spin-dependent Berry curvature. The energy dependence of the spin Hall conductivity can be tuned in order to electrically generate or detect spin currents in spintronic devices.

\subsection{Weyl points on valence bands of AuF$_3$}

\begin{figure*}
	\includegraphics[width=17cm]{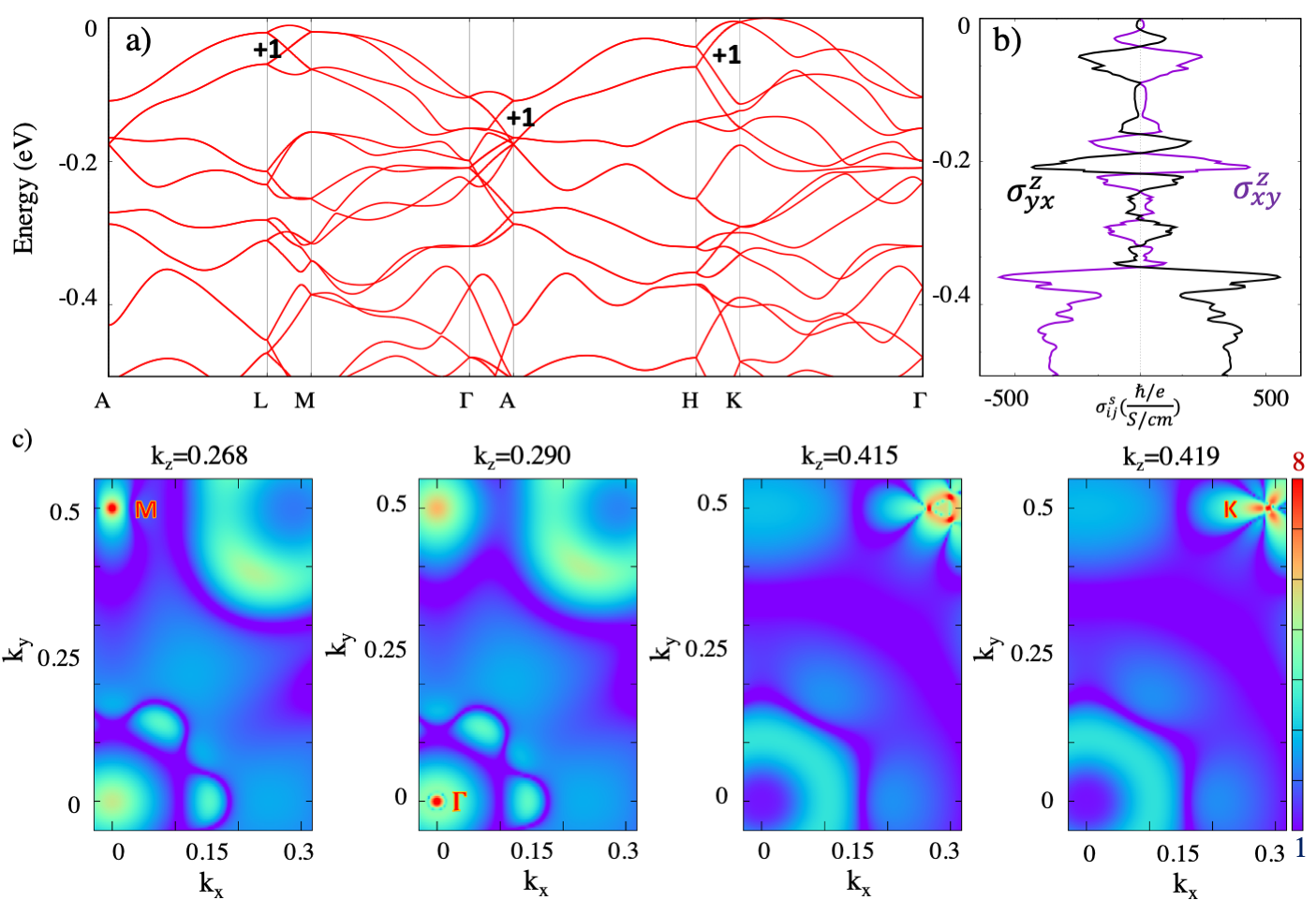}
	\caption{a) Valence band structure for the material AuF$_3$ together with the chirality of the three Weyl points shown between the high-symmetry \textbf{k}-lines (L-M, $\Gamma$-A, H-K) of the BZ. b) $\sigma_{xy}^{z}$ and $\sigma_{yx}^{z}$ components of the spin Hall conductivity tensor as a function of energy. c) $\log(|\Omega_{xy}|+1)$  on different $k_z$-planes (0.268, 0296, 0.415 and 0.419)  where the Weyl points are located in AuF$_3$. Here M, $\Gamma$ and K represent the location of Weyl points at the $k_z$-lines L-M, $\Gamma$-A and H-K respectively. M, K and $\Gamma$ are sources ($+1$ chirality) while the three points close to K are sinks ($-1$ chirality). The Weyl point denoted by $\Gamma$ appears 2 times, the one denoted by M appears 6 times and the one denoted by K appears 4 times. These 12 Weyl points with positive chirality cancel with the 12 Weyl points with negative chirality induced by the three Weyl points close to the one denoted by K. The Fermi level is set to zero.} \label{AuF3-berry}
\end{figure*}

Weyl points were also analysed in the AuF$_3$ material, which crystallizes at the P6$_1$22 space group. The valence band structure for this material is shown in Figure \ref{AuF3-berry}a) where it is also shown the value of the topological charges (chiralities of +1) for the three Weyl points (at the $\Gamma$-A, K-H and M-L lines) generated by the last valence bands. We can notice that the net chirality of these three points is +12 due to the fact that they must be counted twice, four-times and six-times at $\Gamma$-A, K-H and M-L respectively (these are the multiplicities of the 1-cells that appear in Table \ref{Table of isotropy groups}). If only Weyl points are present, extra Weyl points with opposite chirality adding to -12 must appear inside the BZ in order for the total chirality to be zero \cite{Absence-of-neutrinos-in-lattice}. 

In Figure \ref{AuF3-berry}c) we calculated the $xy$-component of the Berry curvature ($\Omega_ {xy}$) at different $k_z$-planes ($k_z$=0.268, 0.296 and 0.419) for the three Weyl points located at the high symmetry lines ($\Gamma$-A, K-H and M-L lines). In Figure \ref {AuF3-berry}c) we can also see three large contributions of the Berry curvature near to the H-point, these points are located at the plane $k_z$=0.415 (in fractional reciprocal lattice vectors).  These three extra Weyl points have each a chirality of $-1$ and they are located on the KHLM-plane. Since the number of equivalent cells (Table \ref{Table of isotropy groups}) to the KHLM-plane is 12, there are 12 Weyl points with negative chirality for a total contribution of $-12$. Therefore the net chirality (total topological charge) of all Weyl points in the BZ vanishes for the AuF$_3$ material as expected.

\subsection{Transport properties of AuF$_3$}

In order to evidence the contribution of the topological states to spin transport properties of AuF$_3$, we have calculated the energy dependence of the spin Hall conductivity as it is shown in Figure \ref {AuF3-berry}b). We can see a large value for the spin Hall conductivity of around 240 ($\hbar$/e)(S/cm) for the energy $E-E_F$$\sim$50 $m$eV, and it could increase to 550 ($\hbar$/e)(S/cm) for $E-E_F$$\sim$360 $m$eV. These values are larger than the one of the pure element Au which is around 40 ($\hbar$/e)(S/cm)  \cite{SpinHallEffects-Sinova}. This result indicates that the F element in the formation of the 4d transition-metal fluorides can increase the spin transport phenomena in pure 4d transition-metal elements.  These findings may help in the quest for the use of Weyl nodes and SOC induced phenomena in developing next-generation of energy-efficient spintronics technology.

\section{Weyl points on conduction bands of A\lowercase{g}F$_3$ and A\lowercase{u}F$_3$}

We have shown in Appendices A and B that the nonsymmorphicity of the space groups P6$_p$ and P6$_p22$ implies the existence of Weyl points along the high symmetry lines $\Gamma$-A, M-L and K-H. One could expect that the number of Weyl points between successive bands is optimal in the sense that the distribution of those Weyl points satisfies the Nielsen-Ninomiya theorem  \cite{Absence-of-neutrinos-in-lattice} of zero net chirality and that the amount of Weyl points is minimal. 

\begin{figure}
\includegraphics[width=8.9cm]{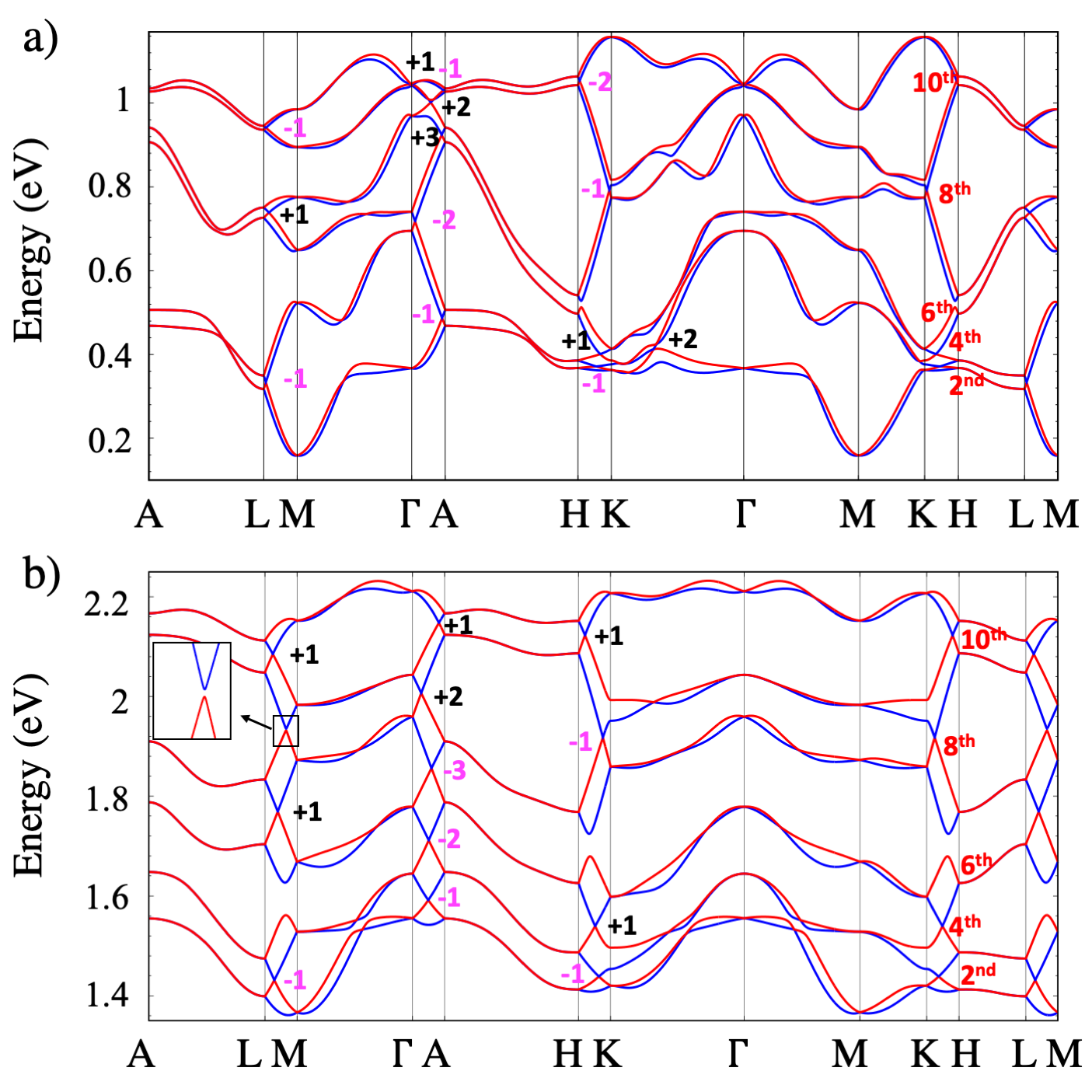}
\caption{a) Lower part of the conduction band structure for AgF$_3$ with the combinatorial structure described in diagram \eqref{complete topological band structure P6122}. b) Lower part of the conduction band structure for AuF$_3$ with identical combinatorial structure as AgF$_3$. Calculated topological chiralities $(\pm 1, \pm 2, \pm3)$ are indicated close to the Weyl points. Numbers indicate the band crossing between $2^{nd}$- $3^{th}$, $4^{th}$- $5^{th}$, $6^{th}$- $7^{th}$, $8^{th}$- $9^{th}$ and $10^{th}$- $11^{th}$ conduction bands.} \label{P6122-bandas} 
\end{figure}

In the case of the space group P6$_122$ we have seen that the electronic bands may assemble as shown in diagram \eqref{complete topological band structure P6122} for the materials AgF$_3$ and AuF$_3$, as shown in Figure \ref{P6122-bandas}.  This assembly incorporates successive bands ($2^{nd}$- $3^{th}$, $4^{th}$- $5^{th}$, $6^{th}$- $7^{th}$, $8^{th}$- $9^{th}$ and $10^{th}$- $11^{th}$) on which there are either three Weyl points or two Weyl points. In the former there is one Weyl point along $\Gamma$-A, one along M-L and one along K-H, and in the latter there is one Weyl point along $\Gamma$-A and the other is either along M-L or along K-H.  Furthermore, the symmetries of the Brillouin zone induce the Weyl points along $\Gamma$-A to appear twice, the Weyl points along K-H four times and the Weyl points along M-L six times (see Table  \ref{Table of isotropy groups}). The distribution of the number of Weyl points together with their chiralities along those three high symmetry lines will allow us to infer certain topological properties of two successive bands. The only tools that we will use to infer these properties are the  Nielsen-Ninomiya theorem \cite{Absence-of-neutrinos-in-lattice}, which tells us that the total chirality is zero, and the counting of the number of times a high symmetry cell appears in the BZ, which is topologically equivalent to the three-dimensional torus.

The specific fact we want to stress is the following. Whenever the chiralities of the Weyl points located in the high symmetry lines
$\Gamma$-A, M-L and  K-H do not add up to a multiple of 12 and no other Weyl points appear in other high symmetry lines, 
then there must exist nodal lines wrapping around high symmetry lines. Furthermore, the chirality of these nodal lines added to the chiralities 
of the Weyl points in the high symmetry lines will always give a multiple of 12. The reason is the following. In the pairs of bands of interest (see Figure \ref{P6122-bandas}) there are no Weyl points at the high-symmetry points in 
reciprocal space. In addition, the multiplicity of the high symmetry planes and the bulk are 12 and 24 respectively (see Table \ref{Table of isotropy groups}). 
As a consequence we have a multiple of 12 for the total chirality of Weyl points outside high symmetry lines and high symmetry points. 
So, if the chiralities along the high symmetry lines $\Gamma$-A, M-L, and K-H are not a multiple of 12 
and there are no other Weyl points in any other high symmetry lines, then it is would be impossible to have total chirality zero with only Weyl points; hence 
there must exist nodal lines or nodal surfaces whose chirality added to the chiralities of the Weyl points in $\Gamma$-A, M-L, and K-H add to a multiple of 12. 

For the P6$_1$22 space group, the high symmetry planes are all fixed only by an antiunitary operator. Since on the pairs of bands of interest (see Figure \ref{P6122-bandas})
there are no degeneracies along the high symmetry points, there will be no degeneracies along any high symmetry plane. Therefore
the presence of nodal surfaces could be disregarded and what will be present are nodal lines.
 We remark here that on the complementary pairs of bands there are indeed degeneracies along high symmetry planes. These pairs of bands have been studied in \cite{B,C,D, Sanchez2019, Rao2019, PhysRevLett.122.076402, Schroter2019} where not only
 appear degeneracies along high symmetry planes but also Kramer-Weyl points on the high symmetry points.

Now, since the chiralities of some of these nodal lines need to add up to a multiple of 12 with the chiralities of the high symmetry lines
$\Gamma$-A, M-L and  K-H, it implies that they must wrap around high symmetry lines. A nodal line
that wraps around the high symmetry lines $\Gamma$-A or  K-H will appear 2 or 4 times respectively, and if it wraps around any other high symmetry line it will appear 6 times.
It is important to highlight here that these nodal lines are induced by the chiralities along the high symmetry lines but its precise location cannot be deduced by the topological analysis. We emphasize that these nodal lines are not protected by glide or mirror symmetries as presented in \cite{Topological-nodal, PhysRevB.94.155108,PhysRevLett.115.036806,PhysRevLett.115.036807}. These nodal lines are not protected by any specific symmetry, they are induced by the complete symmetry group and the explicit distribution of the Weyl points and their chiralities
along high symmetry lines.

We have seen above the case of the third and second to last valence bands of AgF$_3$ and AuF$_3$. For AgF$_3$ the chiralities of the points along the 
$\Gamma$-A, M-L or  K-H add up to zero and there are no other Weyl points (see Table \ref{Weyl points conduction bands AgF3}). For AuF$_3$ the chiralities of the points along the 
$\Gamma$-A, M-L or  K-H add up to 12 and there is another Weyl point whose chirality is $-1$ and its multiplicity is 12 (see Table \ref{Weyl points conduction bands AuF3}).

In Tables \ref{Weyl points conduction bands AgF3} and \ref{Weyl points conduction bands AuF3} we show
the distribution of Weyl points as well as their chiralities on the conduction bands of AgF$_3$ and AuF$_3$ respectively (see Figure \ref{P6122-bandas}). It is worth pointing out the wide range of chirality distributions of the topologically protected Weyl points along the high symmetry lines $\Gamma$-A, M-L and  K-H for each pair of bands. We see that along $\Gamma$-A the Weyl points have chiralities $\pm 1, \pm 2, \pm 3, \pm 2 \pm 1$ for the pairs of bands
(2$^{nd}$,3$^{rd}$), (4$^{th}$,5$^{th}$), (6$^{th}$,7$^{th}$), (8$^{th}$,9$^{th}$) and (10$^{th}$,11$^{th}$) respectively, along M-L the chiralities are always $\pm 1$ and along K-H the chiralities are $\pm 1$ or  $\pm 2$. Also it is important to notice that in all pairs of bands, besides the pair (10$^{th}$,11$^{th}$) of AgF$_3$, the chirality of the Weyl points along $\Gamma$-A, M-L and  K-H add up to a multiple of twelve and there is only presence of Weyl points. 

\begin{table} 
$$\begin{footnotesize}
 \begin{array}{cccccccc} 
 \hline \hline
Location &  \multicolumn{3}{c}{Coordinates \, (2\pi/a_i)} & E-E_F &  \chi & Multi- & Tot. \\
in \; BZ    & k_x & k_y  & k_z  & (eV) &    & plicity&    \\ [0.5ex] 
  \hline \hline
 \multicolumn{8}{c}{Second-Third \; to \; last \; valence \; bands} \\
  \Gamma A & 0 & 0 & 0.449 & -0.220 & +1 & 2 & +2\\
  KH & 0.333 & 0.333 & 0.414 & -0.278 & -2 & 4 & -8 \\
  ML & 0 & 0.500 & 0.463 & -0.267 & +1 & 6 & +6 \\ 
  \hline \hline
  \multicolumn{8}{c}{Second-Third \; conduction \; bands} \\
  \Gamma A & 0 & 0 & 0.454 & 0.487 & -1 & 2 & -2\\
  KH & 0.333 & 0.333 & 0.227 & 0.372 & -1 & 4 & -4 \\
  ML & 0 & 0.500 & 0.468 & 0.333 & -1 & 6 & -6 \\ 
  \Gamma K      & 0.227 & 0.227 & 0 & 0.412 & +2 & 6 & +12 \\
  \Gamma K M & 0.286 & 0.306 & 0 & 0.368 & -1 & 12 & -12 \\
  \Gamma KHA & 0.313 & 0.313 & 0.271 & 0.372 & +1 & 12 & +12\\
  \Gamma KHA & 0.227 & 0.227 & 0.473 & 0.421 & +1 & 12 & +12\\
  \Gamma MLA & 0 & 0.325 & 0.496 & 0.434 & -1 & 12 & -12\\
  \hline
   \multicolumn{8}{c}{Fourth-Fifth   \; \ conduction \; \ bands} \\
   \Gamma A    & 0 & 0 & 0.049 & 0.718  & -2 & 2 & -4\\
   KH                 & 0.333 & 0.333 & 0.165   &  0.401  &  +1 & 4 & +4\\
  KMLH           & 0.325 & 0.496 & 0.089 &  0.457  & -1 & 12 & -12\\
  \Gamma MK & 0.145 & 0.170 & 0       &  0.563  & +1 & 12 & +12\\
  Bulk & 0.180 & 0.205 & 0.183         &  0.501  &-1 & 24 & -24\\
  Bulk  & 0.305    & 0.207    & 0.077   & 0.457  & +1 & 24 & +24\\
  \hline
  \multicolumn{8}{c}{Sixth-Seventh  \; \ conduction \; \ bands} \\
   \Gamma A & 0 & 0 & 0.446& 0.923 & +3 &   2  & +6    \\ 
   ML            & 0  & 0.500 & 0.421        &  0.739   & +1 & 6 & +6\\
  \Gamma KHA           & 0.233 & 0.233  &    0.454     &  0.589  &-1 & 12 & -12\\
   KMLH            & 0.112  & 0.442  & 0.427        &  0.675  &+1 & 12 & +12\\
  \Gamma ALM   & 0 & 0.304  & 0.418  &  0.697  &+1 & 12 & +12\\
  Bulk                & 0.008  & 0.321      & 0.322  &  0.702  &-1 & 24 & -24\\
     \hline 
  \multicolumn{8}{c}{Eighth-Ninth   \; \ conduction \; \ bands} \\
\Gamma A & 0 & 0 & 0.283 & 1.001 & +2 & 2 & +4\\
KH & 0.333 & 0.333 & 0.029 & 0.792 & -1 & 4 & -4 \\
\Gamma KHA & 0.181 & 0.181 & 0.107 & 0.789 & -1 & 12 & -12\\
KMLH & 0.147 & 0.426 & 0.037 & 0.816 & +1 & 12 & +12\\
\hline 
\multicolumn{8}{c}{Tenth-Eleventh  \; \ conduction \; \ bands} \\
\Gamma A & 0 & 0 & 0.471 & 1.040 & -1 & 2 & -2\\
Nodal \ line  & Wrap & \Gamma A & 0.45 & 1.041 & +1 & 2 & +2\\
\Gamma A & 0 & 0 & 0.020 & 1.045 & +1 & 2 & +2\\
KH & 0.333 & 0.333 & 0.473 & 1.033 & -2 & 4 & -8\\
ML & 0 & 0.500 & 0.467 & 0.941 & -1 & 6 & -6 \\
KMLH & 0.083 & 0.457 & 0.459 & 0.979 & +1 & 12 & +12 \\

\hline \hline
\end{array}
\end{footnotesize}$$
\caption{Distribution of Weyl points and nodal lines for the second-third to last valence bands and for the second-third,  fourth-fifth, sixth-seventh, eighth-ninth and tenth-eleventh conduction bands in AgF$_3$, as indicated in Figure \ref{P6122-bandas}(a). Location denotes the high-symmetry line, plane or three-dimensional volume where the Weyl points and nodal lines are located in the Brillouin zone (see Figure \ref{Brillouin}b) and  Table \ref{Table of isotropy groups}), the momentum coordinates $k_x, k_y, k_z$ are given in fractions of the hexagonal reciprocal coordinates, $E-E_F$ denotes the energy respect to Fermi level (in eV), $\chi$ the chirality charge and \textit{Tot.} the total contribution in chirality} \label{Weyl points conduction bands AgF3}
\end{table}

\begin{table} 
$$\begin{footnotesize}
 \begin{array}{cccccccc} 
 \hline \hline
Location &  \multicolumn{3}{c}{Coordinates \, (2\pi/a_i)} & E-E_F &  \chi & Multi- & Tot. \\
in \; BZ    & k_x & k_y  & k_z  & (eV) &    & plicity&    \\ [0.5ex] 
  \hline \hline
 \multicolumn{8}{c}{Second-Third \; to \; last \; valence \; bands} \\
  \Gamma A & 0 & 0 & 0.268 & -0.141 & +1 & 2 & +2\\
  KH & 0.333 & 0.333 & 0.296 & -0.052 & +1 & 4 & +4 \\
  ML & 0.000 & 0.500 & 0.419 & -0.043 & +1 & 6 & +6 \\ 
  KHLM     & 0.345 & 0.309 & 0.415 & -0.052 & -1 & 12 & -12 \\
  \hline \hline
  \multicolumn{8}{c}{Second-Third \; conduction \; bands} \\
  \Gamma A & 0 & 0 & 0.244 & 1.591 & -1 & 2 & -2\\
  KH & 0.333 & 0.333 & 0.182 & 1.438 & -1 & 4 & -4 \\
  ML & 0.000 & 0.500 & 0.353 & 1.431 & -1 & 6 & -6 \\ 
  \Gamma K HA     & 0.294 & 0.294 & 0.212 & 1.441 & +1 & 12 & +12 \\
  \hline
   \multicolumn{8}{c}{Fourth-Fifth   \; \ conduction \; \ bands} \\
  KH                 & 0.333 & 0.333 & 0.241   &  1.539  &  +1 & 4 & +4\\
\Gamma A                & 0 & 0 & 0.251   &  1.711 &  -2 & 2 & -4\\
 \Gamma AHK & 0.251 & 0.251 & 0.236         &  1.554  &+1 & 12 &+ 12\\
  KHLM  & 0.634 & -0.270& 0.237& 1.543&  -1 & 12 & -12\\
   \hline
  \multicolumn{8}{c}{Sixth-Seventh  \; \ conduction \; \ bands} \\
 \Gamma A          & 0  & 0 & 0.293        &  1.833  &-3 & 6 & -6\\ 
   ML            & 0  & 0.500 & 0.300        &  1.768  &+1 & 6 & +6\\  
  \Gamma KHA           & 0.208 & 0.208  &    0.315     &  1.728  &-1 & 12 & -12\\
   \Gamma MLA            & 0  & 0.440  & 0.298        &  1.766 &+1 & 12 & +12\\ 
     \hline 
  \multicolumn{8}{c}{Eighth-Ninth   \; \ conduction \; \ bands} \\
\Gamma A & 0 & 0 & 0.134 & 2.006 & +2 & 2 & +4\\
KH & 0.333 & 0.333 & 0.120 & 1.917 & -1 & 4 & -4 \\
\Gamma KHA & 0.170 & 0.170 & 0.131 & 1.946 & -1 & 12 & -12\\
KHLM & 0.049 & 0.475 & 0.165 & 1.932 & +1 & 12 & +12 \\
\hline 
\multicolumn{8}{c}{Tenth-Eleventh  \; \ conduction \; \ bands} \\
\Gamma A & 0 & 0 & 0.409 & 2.147 & +1 & 2 & +2\\
ML & 0 & 0.5 & 0.376 & 2.084 & +1 & 6 & +6 \\
KH & 0.333 & 0.333 & 0.399 & 2.123 & +1 & 6 & +4 \\
KHLM   & 0.285 & 0.356 & 0.399 & 2.121 & -1 & 12 & -12\\
\hline \hline
\end{array}
\end{footnotesize}$$
\caption{Distribution of Weyl points  for the second-third to last valence bands and for the second-third,  fourth-fifth, sixth-seventh, eighth-ninth and tenth-eleventh conduction bands in AuF$_3$, as indicated in Figure \ref{P6122-bandas}(b). Location denotes the high-symmetry line, plane or three-dimensional volume where the Weyl points are located in the Brillouin zone (see Figure \ref{Brillouin}b) and  Table \ref{Table of isotropy groups}), the momentum coordinates $k_x, k_y, k_z$ are given in fractions of the hexagonal reciprocal coordinates, $E-E_F$ denotes the energy respect to Fermi level (in eV), $\chi$ the chirality charge and \textit{Tot.} the total contribution in chirality} \label{Weyl points conduction bands AuF3}
\end{table}

\begin{figure}
	\includegraphics[width=8.7cm]{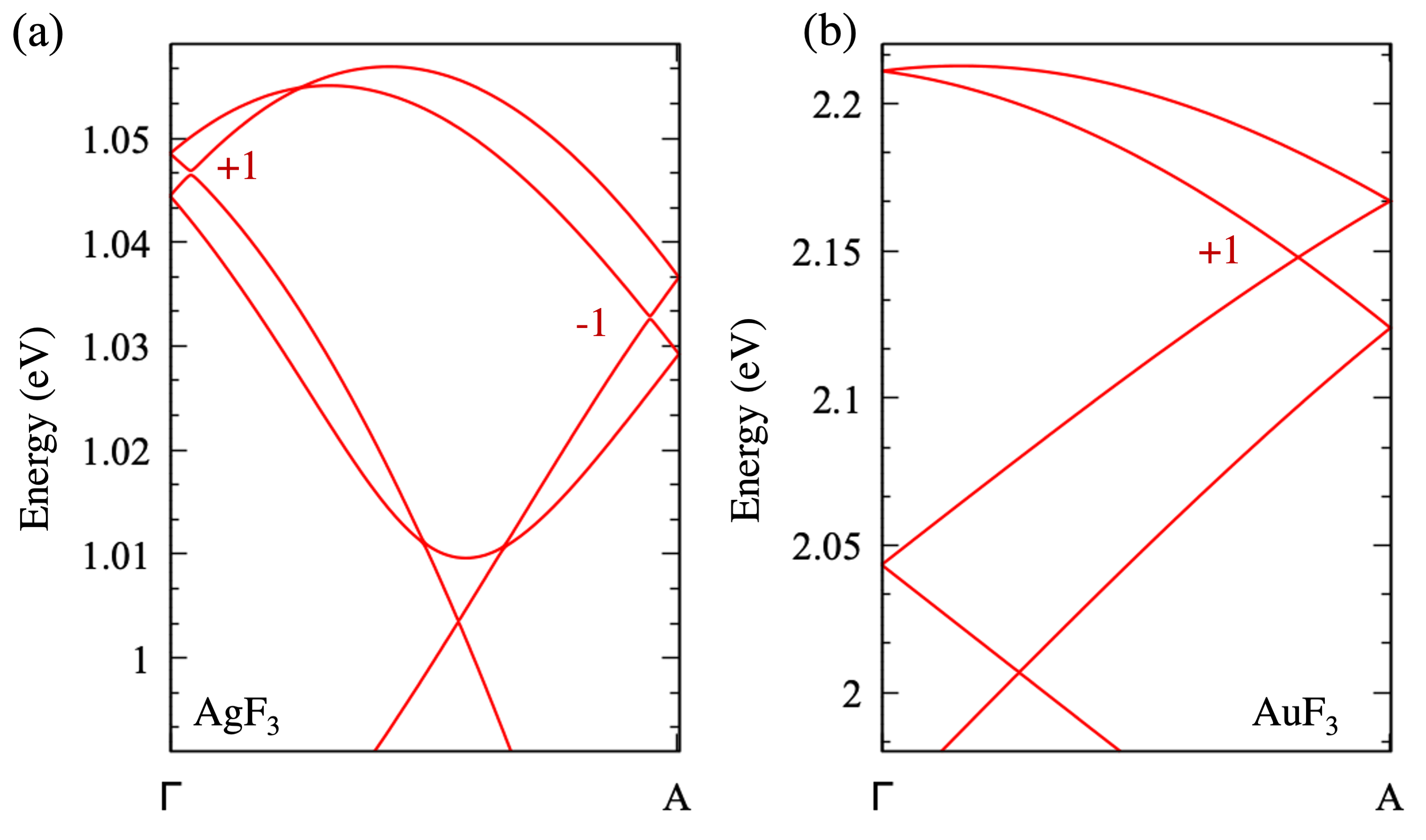}
	\caption{Electronic band structure for a) AgF$_3$ and b) AuF$_3$  for the
	8$^{th}$, 9$^{th}$, 10$^{th}$, 11$^{th}$, and 12$^{th}$ conduction bands along the $\Gamma$-A path. Both preserve the combinatorical accordion-like form
	described in diagram  \eqref{GammaDeltaA P61} but the presentation is different. For AuF$_3$ the presentation is exactly as 
	described in diagram  \eqref{GammaDeltaA P61} and therefore there is only one nodal point for the 10$^{th}$ and 11$^{th}$ conduction bands.
	For AgF$_3$ the upper two vertices of the accordion-like shape have less energy than the second to top ones, thus producing
	two nodal points for the 10$^{th}$ and 11$^{th}$ conduction bands instead of one. This extra nodal point gives rise to the presence
	of a nodal line wrapping around $\Gamma$-A.} \label{GApath}
\end{figure}

\begin{figure}
	\includegraphics[width=8.7cm]{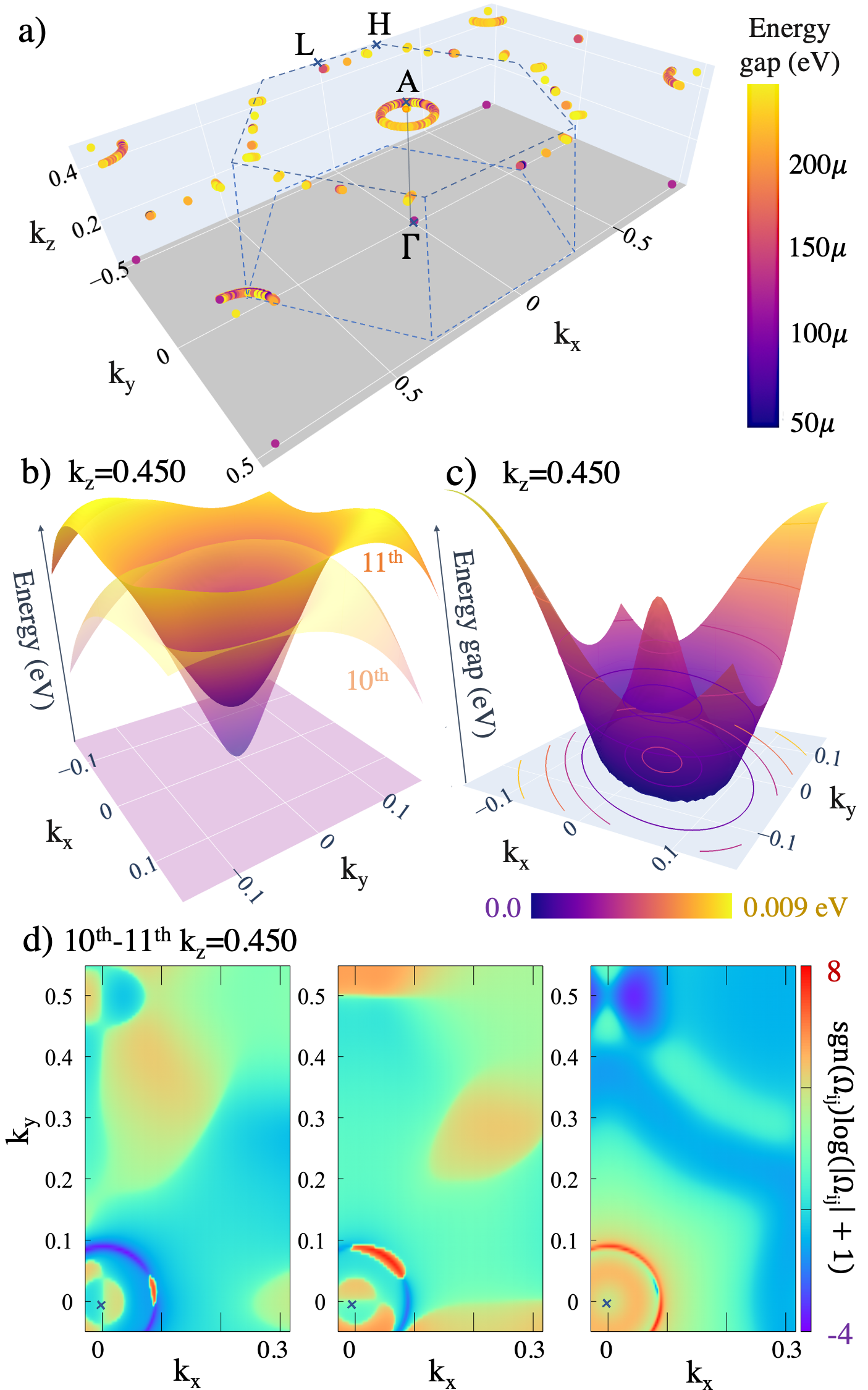}
	\caption{Nodal line wrapping around $\Gamma$-A for the 10$^{th}$ and 11$^{th}$ conduction
	bands in AgF$_3$. a) Position of the $k$-points in fractional reciprocal coordinates whose energy gap is less than $0.00025$ eV. The data for
	this graph comes from WannierTools \cite{wanniertools}. b) Energies and c) energy gap for the two bands localized at the $k_z=0.45$ plane. The data for these two graphs comes directly from DFT calculations. d) $yz$, $xz$ and $xy$ components of the Berry curvature ($\Omega$) localized at $k_z=0.45$.} \label{AgF3-berry+10} 
\end{figure}

The presence of a nodal line wrapping around $\Gamma$-A  on the pair of bands
(10$^{th}$,11$^{th}$) of AgF$_3$ is worth a closer look. 
 Along $\Gamma$-A there are 2 Weyl points with opposite chirality, along M-L there is a Weyl point with chirality $-1$
and along K-H there is a Weyl point of chirality $-2$. The chirality of the points along the high symmetry lines is
 $+2-2-6-8=-14$, and not being a multiple of 12, the existence of nodal lines is enforced.  We have found a nodal line wrapping around the $\Gamma$-A path with chirality $+1$ and a Weyl point of multiplicity 12 on the high symmetry plane KHML with chirality $+1$ thus making the total chirality zero. The information of the Weyl nodal-line can be seen in Table \ref{Weyl points conduction bands AgF3} as well as in Figure \ref{GApath} and  in Figure 
 \ref{AgF3-berry+10}. In Figure \ref{GApath}  we see the energy distribution of the 8$^{th}$, 9$^{th}$, 10$^{th}$, 11$^{th}$ and 12$^{th}$ 
conduction bands of AgF$_3$ and AuF$_3$ along $\Gamma$-A. The distribution of the energy bands for AuF$_3$ presented in Figure \ref{GApath} b) fits the accordion-like presentation
shown in diagram \eqref{GammaDeltaA P61} and therefore there is only one point of intersection of the 10$^{th}$ and 11$^{th}$ conduction bands. For AgF$_3$  the combinatorial description of  diagram \eqref{GammaDeltaA P61} is also preserved, but its presentation is different as it can be seen in Figure \ref{GApath} a). For AgF$_3$ the top two points of the
accordion-like figure have less energy than the second 
 points from top to bottom of the same accordion-like  figure. This fact changes the presentation of the accordion-like shape and produces two intersections of the 10$^{th}$ and 11$^{th}$ conduction bands instead of only one as it is shown in Figure \ref{GApath} b).  This fact explains why the chirality along the high symmetry lines does not
add to a multiple of 12 and therefore a nodal line must be present. The explicit combinatorial distribution of the energy bands of Figure \ref{GApath} a) also shows that the nodal line is stable. The two crossings of the 10$^{th}$ and 11$^{th}$ bands
are topologically enforced and therefore a nodal line with non zero chirality is bound to exist. 

In Figure \ref{AgF3-berry+10} we present evidence on the existence of this nodal line. Figure \ref{AgF3-berry+10}a)
is a 3D plot of the $k$-points in reciprocal space (fractional coordinates) on which there is an energy gap of less than $0.00025$ eV between the 10$^{th}$ and 11$^{th}$ bands. The nodal line around $\Gamma$-A
is clearly visible from this graph and it is located at $k_z=0.45$. Figures \ref{AgF3-berry+10}b) and \ref{AgF3-berry+10}c) represent  respectively the energies of the
10$^{th}$ and 11$^{th}$ conduction bands and the energy gap between the two, both localized at $k_z=0.45$ plane. The energies of Figures  \ref{AgF3-berry+10}b) and \ref{AgF3-berry+10}c)  are obtained by solving the DFT equation directly; the nodal line is also clearly visible. Figure \ref{AgF3-berry+10}d) shows the components $yz$, $zx$ and $xy$ of the Berry curvature
localized at $k_z=0.45$ where the shape of the nodal line could be observed. From the energy dispersion on the $k_x$ and $k_y$ directions
observed in Figure \ref{AgF3-berry+10}c) we evidence that the chirality of the nodal line is $ \pm 1$. From the analysis of the distribution
of all nodal points presented in Table \ref{Weyl points conduction bands AgF3} we infer that the chirality of the nodal line is $+1$.

We found that the resulting band-crossing nodal line in the hexagonal structure of AgF$_3$ cannot be gapped out without breaking any symmetry. This nodal line carries a non-trivial topological charge necessary for Nielsen-Ninomiya to hold and therefore it cannot be removed even by large deformations of the Hamiltonian \cite{Topological-nodal-line-semimetals}.  We checked this statement by performing external pressure on the material by shrinking the $a$ and $b$ lattice constants by 2.3\% and we found that the nodal line persists to exist.

We remark that the topological index that prevents a nodal line to be shrunk to a point is present only on systems whose eigenvectors are real-valued. This happens for instance when time-reversal symmetry and inversion symmetry are preserved and there is no spin-orbit coupling. On these cases the eigenvectors of the Hamiltonian are real-valued and the topological index measures the first Stiefel-Whitney number of the real line bundle defined by one of the eigenstates on a circle that links the nodal line. If the Stiefel-Whitney number is not zero, the line bundle is not trivial and therefore the nodal line cannot be shrunk to a point. This argument appears in \cite[\S 2.2]{Topological-nodal-line-semimetals} and the references therein, although the Stiefel-Whitney classes are never mentioned \footnote{We recommend the classic book
	of Characteristic classes by John Milnor \cite{Milnor} for the interested reader in obstruction theory for vector bundles. Both the Condensed Matter Physics
	and the Mathematics communities would benefit greatly if the appropriate connections with classic constructions in mathematics would be done.}. 
In the case that interests us there is no inversion symmetry. Therefore, all eigenvectors are complex-valued and the topological index vanishes since all complex line bundles over the circle are trivializable. Hence there is no obstruction to shrink nodal lines to Weyl points

Let us finish by noting that the Weyl points do not necessarily appear in pairs of opposite chirality. In materials with nonsymmorphic hexagonal symmetry with inversion symmetry breaking, the distribution of Weyl points must obey the Nielsen-Ninomiya theorem and also must be counted as many times as the multiplicity of the cell on which they appear in the reciprocal space. These conditions break the symmetry under which Weyl points always come in opposite chirality pairs. In this work, we have shown many instances on which this is the case. This feature has been also noted in \cite{Symmetry-Protected-Topological-Triangular-Weyl-Complex} where the topological charge of Weyl phonons in nonsymmorphic trigonal and hexagonal materials have been studied.

\section{Conclusions}

In summary, we used topological corepresentation theory and connectivity group theory to predict Weyl points in nonsymmorphic hexagonal crystal structures.  Weyl points appear at the high-symmetry lines with band crossings in accordion-like and hourglass-like dispersion relations.  Both topology band analysis for hexagonal systems and first-principles calculations for real materials were used to corroborate the behaviour of the energy bands and the location of the Weyl points at the first Brillouin zone. These calculations show a complete agreement. 

We also analysed the distribution of Weyl points in AgF$_3$ and AuF$_3$ materials. Taking into account that the number of times that a Weyl point appears on the Brillouin zone is precisely the multiplicity of the cell on which it appears, together with the Nielsen-Ninomiya theorem, we find interesting distributions of the Weyl points and nodal lines in the valence and conduction bands of these materials. In particular, we show that the Weyl points in AgF$_3$ and AuF$_3$ materials do not appear in pairs of opposite chirality, and moreover, we show the existence
of a Weyl nodal line in a pair of bands where the distribution of Weyl points along the high symmetry lines is of certain kind.

Our results show the simultaneous formation of Weyl points and nodal-lines in 4$d$ transition-metal trifluorides such as AgF$_3$ and AuF$_3$. These materials are feasible Weyl-semimetal candidates with inversion symmetry breaking, with Weyl points protected by 6-fold screw and time-reversal symmetry. In addition, AgF$_3$ and AuF$_3$ exhibit a large intrinsic spin Hall effect (SHE) mainly due to the strong SOC interaction and the particular contribution of the Weyl points and nodal lines in the reciprocal space. These findings may help in the quest for the utilization of Weyl points in developing next-generation of energy-efficient spin-based information technology.

\section{Acknowledgements}
The first author gratefully acknowledges the computing time granted on the supercomputer Mogon at Johannes Gutenberg University Mainz (hpc.uni-mainz.de). The second author thanks the German Service of Academic Exchange (DAAD) for its continuous support. The third author acknowledges the support of the Max Planck Institute for Mathematics in Bonn, Germany. The first and the third authors thank the continuous support of the Alexander Von Humboldt Foundation, Germany. The authors would like to thank the referees for the constructive comments and for pointing out some misconceptions in a preliminary version of this manuscript.

\appendix
\numberwithin{equation}{section}
\renewcommand\theequation{A.\arabic{equation}}

\section{Appendix A\\ Topological band analysys for the P6$_p$ and P6$_p$22 groups}
\label{appendix:a}

In this section we study the band connectivity properties of the symmetry groups P6$_p$ and P6$_p$22
in the presence of spin orbit coupling and the time reversal operator.
In both cases the interesting features of the bands appear in the high symmetry lines 
$\Gamma$-A, K-H and M-L and in the high symmetry plane AKH.
We find the irreducible corepresentations at the high symmetry points  and we study the behaviour
of the eigenfunctions of the screw rotation operator along the fixed high symmetry lines.
This procedure allows us to obtain the topological band connectivity for the symmetry group and allows us in some
cases to produce complete combinatorial band structures that appear in the electronic band structure of
materials with those symmetries.

\subsection{Symmetry groups}

Let us start by recalling the hexagonal materials with screw rotation symmetries. For this consider the symmetries:
\begin{small} 
\begin{align}
Q_{6,p}: (x,y,z) \mapsto & (y, -x+y, z+ {\scriptstyle \frac{p}{6}}) \nonumber  \\
Q_{\overline{6},p}: (x,y,z) \mapsto & (x-y, x, z+ {\scriptstyle \frac{p}{6}}) \nonumber \\
C_3: (x,y,z) \mapsto (-x+y, -x, z),& \ C_2: (x,y,z) \mapsto (-x, -y, z) \nonumber  \\
M_{r}: (x,y,z) \mapsto &(y,x,-z +2r)
\end{align}
\end{small}
together with the translational symmetries of the axis $T_{\hat{x}}, T_{\hat{y}}, T_{\hat{z}}$
whenever the canonical vectors on the $\hat{x}$ and the $\hat{y}$ axis define an hexagonal lattice: i.e.
$\hat{x}=\langle 1,0,0 \rangle$, $\hat{y}= \langle 1/2, \sqrt{3}/2, 0 \rangle$ and $\hat{z}= \langle 0,0,1 \rangle$ in
rectangular coordinates.

The groups of interest of this work are the ones that are generated by the translational symmetries and combinations of the above-defined symmetries. The space groups, the generators, and the relations are summarised the following list (commuting relations are not listed).
We refer to \cite{The-mathematical-theory-of-symmetry-in-solids, BANDREP1, BANDREP2, BANDREP3} and the references therein for a more detailed explanation on the group theory
content of this section.
\begin{small}
\begin{align}
 P6_1 (\#169) :& Q_{\overline{6},1} \label{list of generators in space coordinates}\\
    P6_5 (\#170) :& Q_{6,1} \nonumber\\
    P6_2 (\#171) :& Q_{6,4}, C_2 \nonumber \\
    P6_4 (\#172) :& Q_{6,2}, C_2 \nonumber\\
    P6_3 (\#173) :& Q_{6,3}, C_3 \nonumber\\
        P6_122 (\#178) :& Q_{\overline{6},1}, M_{\scriptstyle \frac{1}{6}}, M_{\scriptstyle \frac{1}{6}} Q_{\overline{6},1} M_{\scriptstyle \frac{1}{6}} = Q_{\overline{6},1}^{-1} \nonumber\\
    P6_522 (\#179) :& Q_{6,1}, M_{\scriptstyle \frac{1}{3}}, M_{\scriptstyle \frac{1}{3}} Q_{6,1} M_{\scriptstyle \frac{1}{3}} =Q_{6,1}^{-1}\nonumber\\
    P6_222 (\#180) :& Q_{6,4}, C_2, M_{\scriptstyle \frac{1}{3}}, M_{\scriptstyle \frac{1}{3}} Q_{6,4} M_{\scriptstyle \frac{1}{3}} =Q_{6,4}^{-1} \nonumber\\
    P6_422 (\#181) :& Q_{6,2}, C_2, M_{\scriptstyle \frac{1}{6}}, M_{\scriptstyle \frac{1}{6}} Q_{6,2} M_{\scriptstyle \frac{1}{6}} =Q_{6,2}^{-1}\nonumber \\
    P6_322 (\#182) :& Q_{6,3}, C_3, M_{0}, M_{0} Q_{6,3} M_{0} =Q_{6,3}^{-1},\nonumber\\ & M_0 C_3M_0=C_3^{-1}. \nonumber 
\end{align}
\end{small}

\begin{figure}  
\begin{center}
\includegraphics[scale=0.310, angle=0]{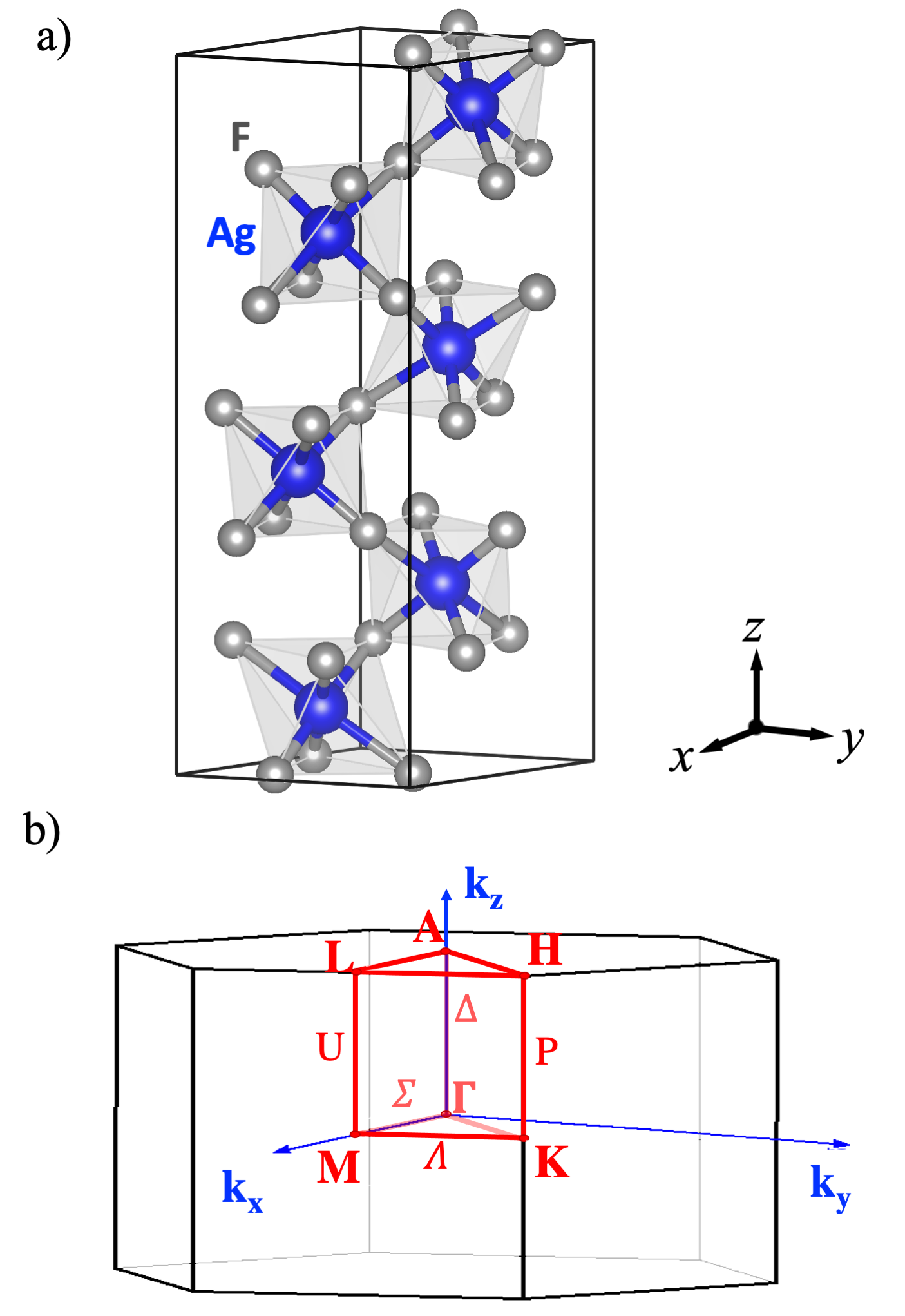}
\caption{a) Crystal structure of the hexagonal phase (P6$_1$22) of AgF$_3$ showing the coordinate system used. The operator $Q_{\overline{6},1}$
rotates $60^\circ$ 
and shifts by 1/6 on the $z$-axis. The operator $M_{\scriptstyle \frac{1}{6}}$ rotates $180^\circ$ around
a line parallel to the $xy$-plane which is perpendicular to the line that joins the bottom atoms of silver.  b) Brillouin zone for the hexagonal Bravais lattice. High-symmetry points (0-cells), high-symmetry lines (1-cells) and high-symmetry planes (2-cells) are indicated according to the Bilbao Crystallographic server \cite{BCS1, BCS2, BCS3}}.
\label{Brillouin}
\end{center}
\end{figure}

The induced action on the momentum space $(k_x,k_y,k_z)$ of all the symmetries $Q_{6,p}$ is simply a six-fold rotation, and the induced action of all the $M_r$'s simply interchanges $k_x$ with $k_y$ and sends $k_z$ to $-k_z$.
The time-reversal operator will be denoted $\mathbb{T}$ and it acts on the momentum coordinates as $\mathbb{T}(k_z,k_y,k_z)=(-k_x,-k_y,-k_z)$. The time-reversal operator acts on electronic states with the property that $\mathbb{T}^2=1$ whenever there is no spin-orbit coupling (NOSOC) or spinless particles, and $\mathbb{T}^2=-1$ whenever there is spin-orbit coupling (SOC) interaction or half-integer spin particles.
In Figure \ref{Brillouin}a) the unit cell for AgF$_3$ with symmetry group P6$_1$22 can be seen. The operator $Q_{\overline{6},1}$
rotates $60^\circ$ 
and shifts by 1/6 on the $z$-axis and the operator $M_{\scriptstyle \frac{1}{6}}$ rotates $180^\circ$ around
a line parallel to the $xy$-plane which is perpendicular to the line that joins the bottom atoms of silver.

The information of the Brillouin zone in momentum space for all these groups appears encoded in Figure \ref{Brillouin}b). Here we have borrowed the notation from the Bilbao Crystallographic Server \cite{BCS1, BCS2, BCS3}. 

Each cell (high-symmetry points, lines and planes) in the Brillouin zone is fixed by certain operators, and the group that these operators define is called the isotropy group of the cell. Some of the topological information of the crystal may be deduced by restricting the attention to the operators $Q$ and $M$. Note that the induced actions of the rotations $C_2$ and $C_3$ on the Brillouin zone are the same as the actions of $Q^3$ and $Q^2$ respectively.  In the table that appears in Table \ref{Table of isotropy groups} we list the subgroups of the isotropy groups generated by these elements localised in specific cells of the Brillouin zone (see Figure \ref{Brillouin}(b)).

\begin{table} 
$$\begin{footnotesize}
 \begin{array}{||c|c|c|c|c||} 
 \hline
 & P6_p &  & P6_p22  &  \\ [0.5ex] 
 \hline
  Cell & Pt. Group & \# cells & Pt. Group  & \# cells \\ [0.5ex] 
  \hline \hline
 \Gamma, A & \langle Q, \mathbb{T} \rangle &1 & \langle Q, M, \mathbb{T} \rangle & 1  \\ 
 \hline
 L, M & \langle Q^3, \mathbb{T} \rangle &3& \langle Q^3, MQ ,\mathbb{T} \rangle & 3\\
 \hline
 H,K &  \langle Q^2,  Q^3\mathbb{T} \rangle &2&   \langle Q^2,M, Q^3\mathbb{T} \rangle  & 2 \\
 \hline \hline
 \Delta (\Gamma A)& \langle Q \rangle & 2& \langle Q, M \mathbb{T} \rangle & 2  \\
 \hline
 U (ML)& \langle Q^3 \rangle &6& \langle Q^3, \mathbb{T}Q^2 M \rangle  & 6\\
 \hline
 P (KH)& \langle Q^2 \rangle & 4& \langle Q^2, \mathbb{T} M Q^3\ \rangle & 4 \\
 \hline
 R (AL), \Sigma (\Gamma M)& \langle Q^3 \mathbb{T} \rangle &  6&
 \langle Q^3 \mathbb{T}, QM \rangle & 6\\
 \hline
 Q (AH), \Lambda (\Gamma K)&  \langle Q^3\mathbb{T} \rangle &6&
 \langle Q^3\mathbb{T}, M \rangle & 6  \\
 \hline
 S (LH), T (MK) &  \langle Q^3\mathbb{T} \rangle & 6&\langle Q^3\mathbb{T}, Q^3M \rangle  & 6 \\
 \hline \hline
 (ALH), (\Gamma MK) &  \langle Q^3 \mathbb{T} \rangle & 6& \langle Q^3 \mathbb{T} \rangle &12    \\
 \hline
 (\Gamma K HA) & \langle \mathrm{Id} \rangle & 12& \langle \mathbb{T} M Q^3 \rangle & 12  \\
 \hline
 (MKHL)& \langle \mathrm{Id} \rangle & 12& \langle M\mathbb{T}  \rangle  & 12 \\
 \hline
 (ALM \Gamma)& \langle \mathrm{Id} \rangle&12& \langle \mathbb{T}Q^2 M \rangle  &12  \\
 \hline \hline
 (ALH\Gamma MK) &\langle \mathrm{Id} \rangle&12& \langle \mathrm{Id} \rangle  &24  \\
 \hline
\end{array}
\end{footnotesize}$$
\caption{Table of subgroups of the isotropy groups which include combinations of the symmetries $Q$ and $\mathbb{T}$
	for P$6_p$ and $Q$, $M$ and $\mathbb{T}$ for P$6_p22$.
	For each cell (high-symmetry points, lines and planes) we list the isotropy group and the number of times that equivalent cells appear in the three-dimensional torus (see Figure \ref{Brillouin}). This last number multiplied by the size of the isotropy group equals $12$ in P$6_p$ and 24 in P$6_p22$.} \label{Table of isotropy groups}
\end{table}

Note that on the 1-cells (high-symmetry lines) $\Gamma$-A and L-M the isotropy groups of the boundary 0-cells (high-symmetry points) are bigger than the ones of the interior of the 1-cell. Also, note that the time-reversal operator fixes the boundary 0-cells and not the interior. This information, together with the fact that the operator $Q$ rotates and translates, induces a very interesting topological structure on the energy bands on these paths. In \cite{PhysRevMaterials.2.074201} the energy bands for the symmetry groups $P6_p$ were studied on the paths $\Gamma$-A and L-M. Here we will extend the study to the groups $P6_p22$ and we will furthermore analyse the band structures for the K-H line and  the other high symmetry lines. To carry out this analysis we need to recall some results in the classification of corepresentations. We will review the classification
scheme of the corepresentations and we will outline the commutation properties of the geometrical operators with the time reversal operator \cite{Wigner}.

\subsection{Corepresentations}
A corepresentation is the name that Wigner \cite[\S 26]{Wigner} assigned when there is a complex representation of a group on which half of its elements act as unitary operators and the other half act as antiunitary operators. Since the time-reversal operator is antiunitary, the isotropy groups of the cells may include antiunitary operators and therefore their representations are corepresentations.

Denote the isotropy group $G$ and let $G_0$ be its subgroup of unitary operators. A corepresentation of $G$ restricts to a complex representation of the group $G_0$ and the classification of the irreducible corepresentations is encoded in the properties of this complex representation. For $u \in G_0$ denote by $\Lambda(u)$ the unitary matrix associated to the chosen representation and denote by $\Lambda$ the representation of $G_0$. Take any antiunitary operator $a_0 \in G$ and define a conjugate complex representation of $G_0$ by the equation

\begin{equation}
\overline{\Lambda}(u) = \Lambda (a_0^{-1} u a_0)^*.
\end{equation}

Now, an irreducible corepresentation of $G$ can be one of three types. The first type contains only one irreducible representation $\Lambda$ of the unitary group $G_0$ and the second type contains twice the irreducible representation $\Lambda \oplus \Lambda$; in these two cases $\Lambda \cong \overline{\Lambda}$.  If $\beta$ is the matrix that transforms $\Lambda$ to $\overline{\Lambda}$ then the associated matrix $M(a_0^2)$ of $a_0^2$ satisfies the equation $\beta \beta^*=\pm M(a_0^2)$. The first type is fulfilled when $\beta \beta^*= M(a_0^2)$
and the second type when $\beta \beta^*=- M(a_0^2)$ \cite[pp. 343]{Wigner}. The third type is when the complex representations $\Lambda$ and $\overline{\Lambda}$ of $G_0$ are inequivalent representations and therefore both must appear in the corepresentation.

\subsection{Antiunitary operators}

Note that the previous information only depends on an antiunitary operator with $\theta^2=\pm 1$. Clearly the time-reversal operator is one of such. Nevertheless, when the time-reversal operator ($\mathbb{T}$) does not belong to the isotropy group of the cell it is important to determine the properties of the other antiunitary symmetries which do belong to the isotropy group.

The antiunitary symmetries that we obtain are of the form $\widehat{F}\mathbb{T}$ where $\widehat{F}$ is the operator on the Hilbert space which lifts the geometrical action given by $F$.  In momentum space the operators $F$ and $\mathbb{T}$ commute, but as operators the commuting relation may be affected by a phase factor.  Let us first study the case when the geometrical action $F$ is the composition $T_{\mathbf{a}}  O_R$ of a rotation $O_R$ by an orthogonal matrix and a translation   $T_{\mathbf{a}}$ by the vector $\mathbf{a}$. Then the operator $\widehat{F}$ equals the composition $\widehat{T}_{\mathbf{a}} \widehat{O}_R$.

The operators that lift rotations $\widehat{O}_R$ commute with the time reversal operator
\begin{equation}
  \widehat{O}_R \mathbb{T}= \mathbb{T} \widehat{O}_R
  \end{equation}
   because there are no non-trivial complex one dimensional representations of the groups $SO(3)$ and $SU(2)$ \cite[Eqn. 26.17]{Wigner}.
On the other hand the commutation relation 
of $\widehat{T}_{\mathbf{a}}$ and $\mathbb{T}$ can be deduced by writing the translation operators in terms of Bloch waves.
Recall that a Bloch wave is of the form
\begin{equation}
\psi_{\mathbf{k}}(\mathbf{r}) = u_{\mathbf{k}}(\mathbf{r}) e^{-i\mathbf{k}\cdot \mathbf{r}}
\end{equation}
where $u_{\mathbf{k}}(\mathbf{r})$ remains invariant under translation by elements of the lattice of the crystal.

Expanding the composition we obtain:
\begin{align} \mathbb{T} \widehat{T}_{\mathbf{a}} \psi_{\mathbf{k}}(\mathbf{r}) = & \mathbb{T}(  u_{\mathbf{k}}(\mathbf{r}+\mathbf{a}) e^{-i\mathbf{k}\cdot \mathbf{r}}e^{-i\mathbf{k}\cdot \mathbf{a}} )  \nonumber\\ 
=&  \overline{u_{-\mathbf{k}}(\mathbf{r}+\mathbf{a})} e^{-i\mathbf{k}\cdot \mathbf{r}}e^{-i\mathbf{k}\cdot \mathbf{a}},\end{align}
and in the opposite order we obtain:
\begin{align}  \widehat{T}_{\mathbf{a}} \mathbb{T}  \psi_{\mathbf{k}}(\mathbf{r}) &= 
 \widehat{T}_{\mathbf{a}} (\overline{u_{-\mathbf{k}}(\mathbf{r})} e^{-i\mathbf{k}\cdot \mathbf{r}} ) \nonumber \\
 &= \overline{u_{-\mathbf{k}}(\mathbf{r}+\mathbf{a})} e^{-i\mathbf{k}\cdot \mathbf{r}}e^{-i\mathbf{k}\cdot \mathbf{a}}.
\end{align}
Hence the operators that lift translations commute with the time reversal operator 
\begin{equation}
\mathbb{T} \widehat{T}_{\mathbf{a}} = \widehat{T}_{\mathbf{a}} \mathbb{T}.
\end{equation}
We emphasise here that the vector $\mathbf{a}$ does not have to belong to the lattice of symmetries of the crystal for the commutativity to hold. Nevertheless whenever $\mathbf{a}$ does belong to the lattice we recover Bloch's theorem $\widehat{T}_{\mathbf{a}} \psi_{\mathbf{k}}(\mathbf{r}) = e^{-i \mathbf{k}\cdot \mathbf{a}}\psi_{\mathbf{k}}(\mathbf{r})$.

Note that the antiunitary operators $\widehat{I} \mathbb{T}$ and
$\widehat{M}_r \mathbb{T}$ both square to 1 on the cells that are fixed by the geometrical operators since either $\widehat{I}$, $\widehat{M}_r$ and $\mathbb{T}$ square to 1 whenever there is no SOC, or $\widehat{I}$, $\widehat{M}_r$ and $\mathbb{T}$ square to -1 whenever there is SOC interaction.

More interestingly, on the cells which are fixed by the operator $\widehat{Q}_{6,p}^3 \mathbb{T}$ we have that
\begin{equation}
(\widehat{Q}_{6,p}^3 \mathbb{T})^2|_{k_z=0} =1
\end{equation} and 
\begin{equation} \label{Q3T on kz pi}
    (\widehat{Q}_{6,p}^3 \mathbb{T})^2|_{k_z=\pi} = \left\{ 
    \begin{array}{cc}
      1   & \mbox{whenever} \  p=2,4\\ 
      -1   &  \mbox{whenever} \ p=1,3,5.
    \end{array} \right.
\end{equation}
since by Bloch's theorem we obtain
\begin{equation}
(\widehat{Q}_{6,p}^3 \mathbb{T})^2=\widehat{Q}_{6,p}^6 \mathbb{T}^2=e^{-i pk_z}
\end{equation}
and we have specialised to the planes $k_z=0$ and $k_z=\pi$.\\
Here is worth pointing out that the antiunitary operator $\widehat{Q}_{6,p}^3 \mathbb{T}$ fixes the points K and H and squares to 1 in K and squares to -1 in H whenever $p$ is odd and squares to 1 whenever $p$ is even. 
Since the path, K-H is fixed by the group generated by $\widehat{Q}_{6,p}^2$
we see that Kramer's degeneracy rule may not occur in all states in K. This fact will be exploited when the topological structure of the bands along the K-H path is analysed.

\section{Topological band analysis}

In this section we will analyse the topological structure of the energy bands along all the high-symmetry lines (1-cells). Of particular interest are the high symmetry 1-cells $\Gamma$-A, M-L, and K-H since the screw rotation operator $\widehat{Q}_{6,p}$  endows them with a very interesting form. 
In \cite{Topological-crossings-hexagonal} a comprehensive study for the high symmetry lines $\Gamma$-A and M-L for the symmetry groups P6$_p$ has been carried out. Here
we expand the analysis to the high symmetry line K-H for the symmetry groups P6$_p$ and we carry out a complete topological band analysis for the symmetry groups P6$_p$22. The similarities of
the band analysis on both cases will be highlighted.

We will start with the symmetry groups $P6_p$ and afterwards we will continue with $P6_p22$. We will take only $p=1,2,3$ since the topological structures for $p=4$ and $p=5$ are equivalent to the ones of $p=2$ and $p=1$ respectively.

For each high symmetry line parallel to the six-fold rotation axis and each $p$ we will also reconstruct the combinatorial band diagrams using the information found in the Bilbao Crystallographic Server (BCS) under the BANDREP menu \cite{BANDREP1, BANDREP2, BANDREP3}.
We will write the compatibility relations of interest, we will construct the associated band representation and   we will match this band representation with the one we have constructed. We refer to  \cite{BANDREP1, BANDREP2, BANDREP3} for the explanations on the symbology.

\subsection{Symmetry groups $P6_p$}

\subsubsection{Topological band analysis on $\Gamma$-A}

The isotropy groups of the $\Gamma$ and A points are generated by the elements $Q_{6,p}$ and $\mathbb{T}$ while the isotropy group of the $\Gamma$-A path is generated only by $Q_{6,p}$. The operators $\widehat{Q}_{6,p}$ and $\mathbb{T}$ commute and in the presence of SOC interaction they satisfy the equations 
\begin{equation}\widehat{Q}_{6,p}^6=-e^{-ik_zp} \ \ \mbox{and} \ \mathbb{T}^2=-1.\end{equation}

The operator $\widehat{Q}_{6,p}$ may be diagonalised on this path with eigenvalues
\begin{equation} \label{eigenvectors Q GammaDeltaA line}
\widehat{Q}_{6,p}  \psi_{l}(\mathbf{k}) = e^{i \pi \frac{l}{6}} e^{-ik_z\frac{p}{6}} \psi_{l}(\mathbf{k})
\end{equation}
with $l \in \{1,3,5,7,9,11\}$. Specialising to the points $\Gamma$ and A we obtain the representations
\begin{align}
\widehat{Q}_{6,p}  \psi_{l}(\Gamma) &= e^{i \pi \frac{l}{6}}  \psi_{l}(\Gamma), \nonumber\\ 
 \widehat{Q}_{6,p}  \psi_{l}(A) &= e^{i \pi \frac{l-p}{6}} \psi_{l}(A).
 \end{align}
These representations may be lifted to corepresentations of the groups generated by $\widehat{Q}_{6,p}$ and $\mathbb{T}$ in the following way. On $\Gamma$ the pairs $\{\psi_{1}(\Gamma), \psi_{11}(\Gamma) \}$, $\{\psi_{3}(\Gamma), \psi_{9}(\Gamma) \}$ and $\{\psi_{5}(\Gamma), \psi_{7}(\Gamma) \}$ define irreducible corepresentations of the third type since in this case $\psi_l(\Gamma)$ and $\overline{\psi_l(\Gamma)}=\psi_{12-l}(\Gamma)$ are 
not isomorphic representations. Therefore on $\Gamma$ the states $\psi_l(\Gamma)$ and $\psi_{12-l}(\Gamma)$ have the same energy.

On $A$ the relation on the representations depends on the value of $p$.
For $p=1$ the pairs $\{\psi_{3}(A), \psi_{11}(A) \}$ and $ \{\psi_{5}(A), \psi_{9}(A) \}$ define corepresentations of the third type and both the representations $\psi_{1}(A)$ and $ \psi_{7}(A)$ must appear twice since 
the operator $\mathbb{T}$ forces each of these representations to become quaternionic and therefore they must come in doublets. Hence the corepresentations defined by   $\psi_{1}(A) \oplus \psi_{1}(A)$ and 
$\psi_{7}(A) \oplus \psi_{7}(A)$ are of the second type. In this case, the combinatorial description of the band appears in figure \eqref{GammaDeltaA P61} where the bands are denoted by the eigenfunctions $\psi_j$ and the values on $\Gamma$ and A are the eigenvalues of the eigenfunctions in $\Gamma$ and A respectively:

\begin{equation} \label{GammaDeltaA P61}
\xymatrix@1@R=0.2cm@C=1.5cm{
\Gamma &\Delta& A\\
{e^{i\pi/6} \atop e^{i\pi 11/6}}  \bullet \ar@{-}[rr]^(.5){\psi_1} \ar@{-}[rrd]^(.8){\psi_{11}} &&\bullet {e^0 \atop e^0 }\\
{e^{i\pi /6} \atop e^{i\pi 11/6}} \bullet \ar@{-}[rrd]^(.8){\psi_{11}} \ar@{-}[rru]^(.2){\psi_1} && \bullet {e^{i\pi 10/6} \atop e^{i\pi 2/6}}\\
{e^{i\pi 3/6} \atop e^{i\pi 9/6}} \bullet \ar@{-}[rrd]^(.8){\psi_{9}} \ar@{-}[rru]^(.2){\psi_3} && \bullet {e^{i\pi 10/6} \atop e^{i\pi 2/6}}\\
{e^{i\pi 3/6} \atop e^{i\pi 9/6}} \bullet \ar@{-}[rrd]^(.8){\psi_{9}} \ar@{-}[rru]^(.2){\psi_3} && \bullet {e^{i\pi 8/6} \atop e^{i\pi 4/6}}\\
{e^{i\pi 5/6} \atop e^{i\pi 7/6}} \bullet \ar@{-}[rrd]^(.8){\psi_{7}} \ar@{-}[rru]^(.2){\psi_5} && \bullet {e^{i\pi 8/6} \atop e^{i\pi 4/6}}\\
{e^{i\pi 5/6} \atop e^{i\pi 7/6}} \bullet \ar@{-}[rr]_(.5){\psi_{7}} \ar@{-}[rru]^(.2){\psi_5} && \bullet {e^{i\pi 6/6} \atop e^{i\pi 6/6}}
} \end{equation}

Now, from the Bilbao Crystallographic Server (BCS) we obtain the minimal set of compatibility relations:
\begin{small}
\begin{align*}
    \overline{\Gamma}_9\overline{\Gamma}_{12}(2) & \to \overline{\Delta}_9(1) \oplus \overline{\Delta}_{12}(1) &  2\overline{\Delta}_{12}(2) & \leftarrow \overline{A}_9\overline{A}_9(2) \\
    \overline{\Gamma}_7\overline{\Gamma}_{8}(2) & \to \overline{\Delta}_7(1) \oplus \overline{\Delta}_{8}(1)  &  2\overline{\Delta}_{11}(2) & \leftarrow \overline{A}_{10}\overline{A}_{10}(2) \\
    \overline{\Gamma}_{10}\overline{\Gamma}_{11}(2) & \to \overline{\Delta}_{10}(1) \oplus \overline{\Delta}_{11}(1) & \overline{\Delta}_{7}(1) \oplus \overline{\Delta}_{9}(1)& \leftarrow \overline{A}_{8}\overline{A}_{12}(2) \\
    &&  \overline{\Delta}_{8}(1) \oplus \overline{\Delta}_{10}(1)& \leftarrow \overline{A}_{7}\overline{A}_{11}(2).
\end{align*}
\end{small}
These representations can be assembled into diagram \eqref{GammaDeltaA P61 Bandrep}  agreeing completely with diagram \eqref{GammaDeltaA P61}.
\begin{equation} \label{GammaDeltaA P61 Bandrep}
\xymatrix@1@R=0.3cm@C=1.5cm{
\overline{\Gamma}_{9}\overline{\Gamma}_{12}(2)  \bullet \ar@{-}[rr]^(.5){\overline{\Delta}_{12}(1)} \ar@{-}[rrd]^(.8){\overline{\Delta}_{9}(1)} &&\bullet \overline{A}_{9}\overline{A}_{9}(2)\\
\overline{\Gamma}_{9}\overline{\Gamma}_{12}(2) \bullet \ar@{-}[rrd]^(.8){\overline{\Delta}_{9}(1)} \ar@{-}[rru]^(.2){\overline{\Delta}_{12}(1)} && \bullet \overline{A}_{8}\overline{A}_{12}(2)\\
\overline{\Gamma}_{7}\overline{\Gamma}_{8}(2) \bullet \ar@{-}[rrd]^(.8){\overline{\Delta}_{8}(1)} \ar@{-}[rru]^(.2){\overline{\Delta}_{7}(1)} && \bullet \overline{A}_{8}\overline{A}_{12}(2)\\
\overline{\Gamma}_{7}\overline{\Gamma}_{8}(2) \bullet \ar@{-}[rrd]^(.8){\overline{\Delta}_{8}(1)} \ar@{-}[rru]^(.2){\overline{\Delta}_{7}(1)} && \bullet \overline{A}_{7}\overline{A}_{11}(2)\\
\overline{\Gamma}_{10}\overline{\Gamma}_{11}(2) \bullet \ar@{-}[rrd]^(.8){\overline{\Delta}_{11}(1)} \ar@{-}[rru]^(.2){\overline{\Delta}_{10}(1)} && \bullet \overline{A}_{7}\overline{A}_{11}(2)\\
\overline{\Gamma}_{10}\overline{\Gamma}_{11}(2) \bullet \ar@{-}[rr]_(.5){\overline{\Delta}_{11}(1)} \ar@{-}[rru]^(.2){\overline{\Delta}_{10}(1)} && \bullet \overline{A}_{10}\overline{A}_{10}(2)
}\end{equation}

For $p=2$ we have that the pairs $\{\psi_{1}(A), \psi_{3}(A) \}$, $\{\psi_{5}(A), \psi_{11}(A) \}$ and $\{\psi_{7}(A), \psi_{9}(A) \}$
define corepresentations of the third type and the combinatorial description 
of the band appears in diagram \eqref{GammaDeltaA P62}:
\begin{equation} \label{GammaDeltaA P62}
\xymatrix@1@R=0.2cm@C=1.5cm{
\Gamma &\Delta& A\\
{e^{i\pi/6} \atop e^{i\pi 11/6}}  \bullet \ar@{-}[rr]^(.5){\psi_1} \ar@{-}[rrd]^(.8){\psi_{11}} &&\bullet {e^{i\pi11/6} \atop e^{i\pi 1/6}}\\
{e^{i\pi 3/6} \atop e^{i\pi 9/6}} \bullet \ar@{-}[rrd]^(.8){\psi_{7}} \ar@{-}[rru]^(.2){\psi_3} && \bullet {e^{i\pi 9/6} \atop e^{i\pi 3/6}}\\
{e^{i\pi 5/6} \atop e^{i\pi 7/6}} \bullet \ar@{-}[rr]_(.5){\psi_{7}} \ar@{-}[rru]^(.2){\psi_5} && \bullet {e^{i\pi 7/6} \atop e^{i\pi 5/6}}
}
\end{equation}

In this case the minimal set of compatibility relations from BCS is:
\begin{small}
\begin{align*}
    \overline{\Gamma}_{7}\overline{\Gamma}_{8}(2) & \to \overline{\Delta}_{7}(1) \oplus \overline{\Delta}_{8}(1) &  \overline{\Delta}_{7}(1) \oplus \overline{\Delta}_{12}(1) & \leftarrow \overline{A}_8\overline{A}_9(2) \\
    \overline{\Gamma}_{9}\overline{\Gamma}_{12}(2) & \to \overline{\Delta}_{9}(1) \oplus \overline{\Delta}_{12}(1)  &  \overline{\Delta}_{8}(1) \oplus \overline{\Delta}_{9}(1) & \leftarrow \overline{A}_{10}\overline{A}_{7}(2) \\
    \overline{\Gamma}_{10}\overline{\Gamma}_{11}(2) & \to \overline{\Delta}_{10}(1) \oplus \overline{\Delta}_{11}(1) & \overline{\Delta}_{9}(1) \oplus \overline{\Delta}_{10}(1)& \leftarrow \overline{A}_{11}\overline{A}_{12}(2),
\end{align*}
\end{small}
which leads to the diagram \eqref{GammaDeltaA P62 Bandrep} in agreement with diagram \eqref{GammaDeltaA P62}.
\begin{equation} \label{GammaDeltaA P62 Bandrep}
\xymatrix@1@R=0.3cm@C=1.5cm{
\overline{\Gamma}_{7}\overline{\Gamma}_{8}(2) \bullet \ar@{-}[rr]^(.5){\overline{\Delta}_{7}(1)} \ar@{-}[rrd]^(.8){\overline{\Delta}_{8}(1)} &&\bullet \overline{A}_{8}\overline{A}_{9}(2) \\
\overline{\Gamma}_{12}\overline{\Gamma}_{9}(2)\bullet \ar@{-}[rrd]^(.8){\overline{\Delta}_{9}(1)} \ar@{-}[rru]^(.2){\overline{\Delta}_{12}(1)} && \bullet \overline{A}_{10}\overline{A}_{7}(2) \\
\overline{\Gamma}_{10}\overline{\Gamma}_{11}(2) \bullet \ar@{-}[rr]_(.5){\overline{\Delta}_{10}(1)} \ar@{-}[rru]^(.2){\overline{\Delta}_{11}(1)} && \bullet \overline{A}_{11}\overline{A}_{12}(2) 
}
\end{equation}
Finally, for $p=3$ the pairs $\{\psi_{1}(A), \psi_{5}(A) \}$ and $\{\psi_{7}(A), \psi_{11}(A) \}$ define corepresentations of the third type, and the representations $\psi_{3}(A)$ and $\psi_{9}(A)$ must appear twice thus making $\psi_{3}(A) \oplus \psi_{3}(A)$ and $\psi_{9}(A) \oplus \psi_{9}(A)$ corepresentations of the second type. The combinatorial structure appears in diagram \eqref{GammaDeltaA P63}:
\begin{equation} \label{GammaDeltaA P63}
\xymatrix@1@R=0.2cm@C=1.5cm{
\Gamma &\Delta& A\\
{e^{i\pi/6} \atop e^{i\pi 11/6}}  \bullet \ar@{-}[rr]^(.5){\psi_1} \ar@{-}[rrd]^(.8){\psi_{11}} &&\bullet {e^{i\pi10/6} \atop e^{i\pi 2/6}}\\
{e^{i\pi 5/6} \atop e^{i\pi 7/6}} \bullet \ar@{-}[rr]_(.5){\psi_{7}} \ar@{-}[rru]^(.2){\psi_5} && \bullet {e^{i\pi 8/6} \atop e^{i\pi 4/6}}\\
{e^{i\pi3/6} \atop e^{i\pi 9/6}}  \bullet \ar@{-}[rr]^(.5){\psi_3} \ar@{-}[rrd]^(.8){\psi_{9}} &&\bullet {e^{0} \atop e^{0}}\\
{e^{i\pi 3/6} \atop e^{i\pi 9/6}} \bullet \ar@{-}[rr]_(.5){\psi_{9}} \ar@{-}[rru]^(.2){\psi_3} && \bullet {e^{i\pi 6/6} \atop e^{i\pi 6/6}}
}
\end{equation}
The minimal set of compatibility relations from BCS for this case is:
\begin{small}
\begin{align*}
    \overline{\Gamma}_{7}\overline{\Gamma}_{8}(2) & \to \overline{\Delta}_{7}(1) \oplus \overline{\Delta}_{8}(1) &  \overline{\Delta}_{8}(1) \oplus \overline{\Delta}_{8}(1) & \leftarrow \overline{A}_7\overline{A}_7(2) \\
    \overline{\Gamma}_{9}\overline{\Gamma}_{12}(2) & \to \overline{\Delta}_{9}(1) \oplus \overline{\Delta}_{12}(1)  &  \overline{\Delta}_{7}(1) \oplus \overline{\Delta}_{7}(1) & \leftarrow \overline{A}_{8}\overline{A}_{8}(2) \\
    \overline{\Gamma}_{10}\overline{\Gamma}_{11}(2) & \to \overline{\Delta}_{10}(1) \oplus \overline{\Delta}_{11}(1) & \overline{\Delta}_{10}(1) \oplus \overline{\Delta}_{12}(1)& \leftarrow \overline{A}_{11}\overline{A}_{9}(2) \\
    && \overline{\Delta}_{11}(1) \oplus \overline{\Delta}_{9}(1)& \leftarrow \overline{A}_{10}\overline{A}_{12}(2),
\end{align*}
\end{small}
which leads to diagram \eqref{GammaDeltaA P63 Bandrep} in agreement with diagram \eqref{GammaDeltaA P63}.
\begin{equation} \label{GammaDeltaA P63 Bandrep}
\xymatrix@1@R=0.3cm@C=1.5cm{
\overline{\Gamma}_{9}\overline{\Gamma}_{12}(2)  \bullet \ar@{-}[rr]^(.5){\overline{\Delta}_{9}(1)} \ar@{-}[rrd]^(.8){\overline{\Delta}_{12}(1)} &&\bullet \overline{A}_{10}\overline{A}_{12}(2)\\
\overline{\Gamma}_{10}\overline{\Gamma}_{11}(2)\bullet \ar@{-}[rr]_(.5){\overline{\Delta}_{10}(1)} \ar@{-}[rru]^(.2){\overline{\Delta}_{11}(1)} && \bullet \overline{A}_{11}\overline{A}_{9}(2)\\
 \overline{\Gamma}_{7}\overline{\Gamma}_{8}(2)  \bullet \ar@{-}[rr]^(.5){\overline{\Delta}_{8}(1)} \ar@{-}[rrd]^(.8){\overline{\Delta}_{7}(1)} &&\bullet \overline{A}_7\overline{A}_7(2)\\
 \overline{\Gamma}_{7}\overline{\Gamma}_{8}(2) \bullet \ar@{-}[rr]_(.5){\overline{\Delta}_{7}(1)} \ar@{-}[rru]^(.2){\overline{\Delta}_{8}(1)} && \bullet \overline{A}_8\overline{A}_8(2)
}
\end{equation}

\begin{figure}
\includegraphics[width=8.5cm]{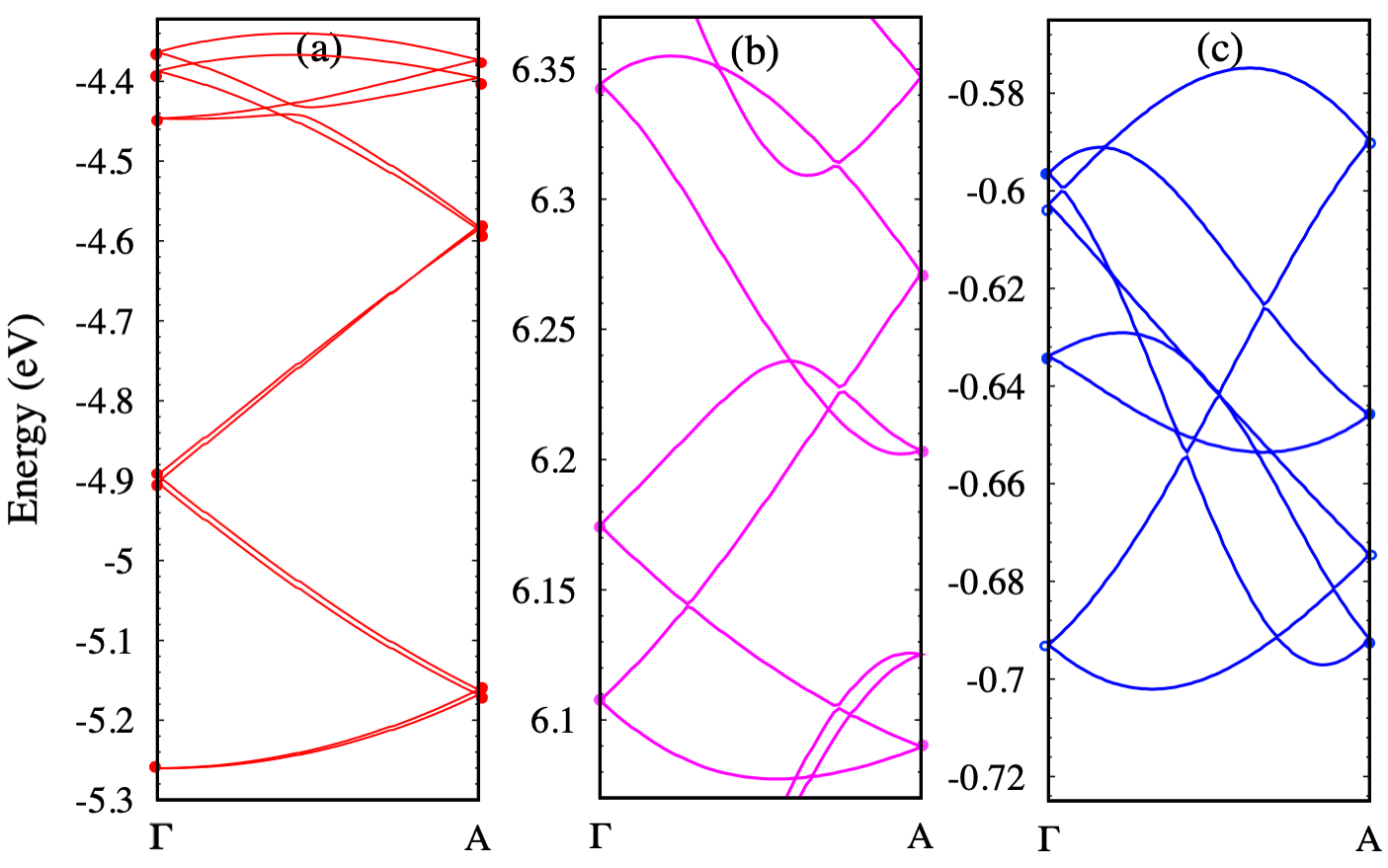}
\caption{Electronic band structure in $\Gamma$-A for the space groups a)P6$_1$ (In$_2$Se$_3$), b) P6$_4$ (KCaNd(PO$_4$)$_2$) and c) P$6_3$ (PI$_3$) space groups. The topological structure of the energy bands match with the ones presented in diagrams 
\eqref{GammaDeltaA P61}, \eqref{GammaDeltaA P62} and \eqref{GammaDeltaA P63} respectively.} \label{bands-GA}
 \end{figure}

Parallel to the previous combinatorial description we have carried out computational first-principles calculations of hexagonal materials with the P$6_1$, P$6_2$ and P$6_3$ space groups in order to check the theoretical predicted topological band structures. We have plotted the electronic band structure along the $\Gamma$-A path for each space groups in Figure \ref{bands-GA} and we can see that the expected combinatorial structures are recovered if we compare them with diagrams \eqref{GammaDeltaA P61}, \eqref{GammaDeltaA P62} and \eqref{GammaDeltaA P63}. As a by-product of our calculations we notice that a set of Weyl points are generated by the accordion-like (\ref{bands-GA}a) and hourglass-like dispersion figures (\ref{bands-GA}b and c).

\subsubsection{Topological band analysis on M-L}

The isotropy groups of the points M and L are generated by $Q^3_{6,p}$ and $\mathbb{T}$ while the isotropy group of the path M-L between the points is only generated by $Q^3_{6,p}$. Since $(\widehat{Q}^3_{6,p})^2= -e^{-ik_zp}$
we may diagonalise this operator on the path M-L having eigenvalues
\begin{equation} \label{eigenvectors Q MUL line}
\widehat{Q}^3_{6,p} \psi_{\pm}(\mathbf{k}) = \pm i e^{-ik_z\frac{p}{2}}\psi_{\pm}(\mathbf{k}).
\end{equation}
Specialising on M and L we obtain the representations
\begin{align}
\widehat{Q}_{6,p}^3  \psi_{\pm}(M) &= \pm i  \psi_{\pm}(M), \nonumber \\
  \widehat{Q}_{6,p}^3  \psi_{\pm}(L) &= \pm i e^{-i \pi \frac{p}{2}} \psi_{\pm}(L).
  \end{align} 

On M the pair $\{\psi_+(M),\psi_-(M) \}$ defines a corepresentation of the  third type. On L we have that for $p=1,3$ the representations
$\psi_+(L)$ and $\psi_-(L)$ must appear twice so that $\psi_+(L) \oplus \psi_+(L)$ and $\psi_-(L) \oplus \psi_-(L)$ define representations of the second type. 
The combinatorial structures from M to L appear in diagram \eqref{MUL P61} for $p=1$ and in diagram \eqref{MUL P63} for $p=3$:

\begin{equation} \label{MUL P61}
\xymatrix@1@R=0.2cm@C=1.5cm{
M & U& L\\
{i \atop -i} \bullet \ar@{-}[rr]^(.5){\psi_+} \ar@{-}[rrd]^(.8){\psi_-} &&\bullet {1 \atop 1 }\\
{i \atop -i}\bullet \ar@{-}[rr]_(.5){\psi_-} \ar@{-}[rru]^(.2){\psi_+} && \bullet {-1 \atop -1 }
}\end{equation}
\begin{equation} \label{MUL P63}
\xymatrix@1@R=0.2cm@C=1.5cm{
{i \atop -i} \bullet \ar@{-}[rr]^(.5){\psi_+} \ar@{-}[rrd]^(.8){\psi_-} &&\bullet {-1 \atop -1 }\\
{i \atop -i}\bullet \ar@{-}[rr]_(.5){\psi_-} \ar@{-}[rru]^(.2){\psi_+} && \bullet {1 \atop 1 }
}\end{equation}

The minimal set of compatibility relations  from BCS for $p=1$ and $p=3$ is:
\begin{small}
\begin{align*}
    \overline{M}_{3}\overline{M}_{4}(2) & \to \overline{U}_{3}(1) \oplus \overline{U}_{4}(1) &  2\overline{U}_{4}(1) & \leftarrow \overline{L}_3\overline{L}_3(2) \\
     &   &  2\overline{U}_{3}(1) & \leftarrow \overline{L}_4\overline{L}_4(2)
\end{align*}
\end{small}
leading to diagram  \eqref{MUL P61 P63 Bandrep}. This is in agreement with diagrams \eqref{MUL P61} and \eqref{MUL P63}.
\begin{equation} \label{MUL P61 P63 Bandrep}
\xymatrix@1@R=0.2cm@C=1.5cm{
\overline{M}_{3}\overline{M}_{4}(2) \bullet \ar@{-}[rr]^(.5){\overline{U}_{4}(1)} \ar@{-}[rrd]^(.8){\overline{U}_{3}(1)} &&\bullet \overline{L}_3\overline{L}_3(2)\\
\overline{M}_{3}\overline{M}_{4}(2)\bullet \ar@{-}[rr]_(.5){\overline{U}_{3}(1)} \ar@{-}[rru]^(.2){\overline{U}_{4}(1)} && \bullet \overline{L}_4\overline{L}_4(2)
}\end{equation}

\begin{figure}
\includegraphics[width=8.5cm]{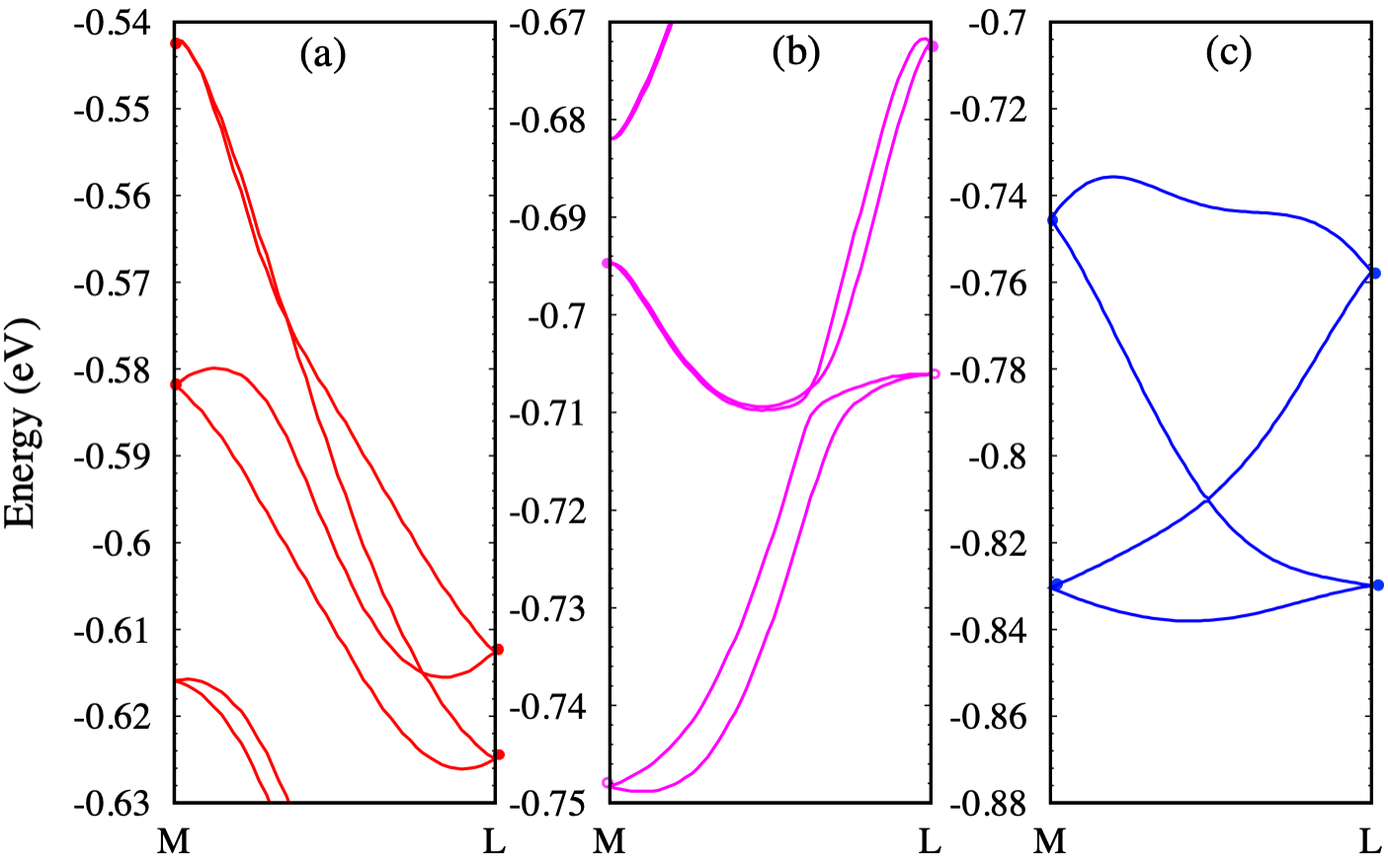}
\caption{Electronic band structure from M-L for a) P6$_1$ (In$_2$Se$_3$), b) P6$_4$ (KCaNd(PO$_4$)$_2$) and c) P$6_3$ (PI$_3$) space groups. The topological structure of the energy bands match the ones presented in diagrams \eqref{MUL P61}, \eqref{MUL P62} and \eqref{MUL P63} respectively.} \label{bands-ML}
\end{figure}

For $p=2$ the pair $\{\psi_+(L),\psi_-(L) \}$ defines a corepresentation of the third type. The combinatorial structure appears in  diagram \eqref{MUL P62}:
\begin{equation} \label{MUL P62}
\xymatrix@1@R=0.2cm@C=1.5cm{
M & U& L\\
{i \atop -i} \bullet \ar@{-}@/^1pc/[rr]_(.7){\psi_+} \ar@{-}@/_1pc/[rr]^(0.3){\psi_-} &&\bullet {-i \atop i }
}\end{equation} 
The minimal set of compatibility relations from BCS for this case is
\begin{small}
\begin{align*}
    \overline{M}_{3}\overline{M}_{4}(2) & \to \overline{U}_{3}(1) \oplus \overline{U}_{4}(1) &  \overline{U}_{3}(1) \oplus \overline{U}_{4}(1) & \leftarrow \overline{L}_3\overline{L}_4(2) 
\end{align*}
\end{small}
which assemble into diagram  \eqref{MUL P62 Bandrep} which is in agreement with diagram \eqref{MUL P62}.
\begin{equation} \label{MUL P62 Bandrep}
\xymatrix@1@R=0.2cm@C=1.5cm{
\overline{M}_{3}\overline{M}_{4}(2) \bullet \ar@{-}@/^1pc/[rr]_(.7){\overline{U}_{3}(1)} \ar@{-}@/_1pc/[rr]^(0.3){\overline{U}_{4}(1)} &&\bullet \overline{L}_3\overline{L}_4(2)
}\end{equation}
Similarly as with the previous analysis for $\Gamma$-A, we have carried out first-principle calculations for hexagonal materials with the P$6_p$ symmetry along the M-L path. The results for the electronic band structures are presented in Figure \ref{bands-ML}. It can be seen that the previously described combinatorial structure of the energy bands is recovered.

\subsubsection{Topological band analysis on K-H}
The isotropy groups of the points K and H are generated by $Q^2_{6,p}$ and $Q^3_{6,p}\mathbb{T}$ while the isotropy group of the path K-H between the points is only generated by $Q^2_{6,p}$. Since $(\widehat{Q}^2_{6,p})^3= -e^{-ik_zp}$
we may diagonalise this operator on the path K-H having eigenvalues
\begin{equation} \label{eigenvectors Q KPH line}
\widehat{Q}^2_{6,p} \psi_{l}(\mathbf{k}) = e^{-i\pi\frac{l}{3}} e^{-ik_z\frac{p}{3}}\psi_{l}(\mathbf{k})
\end{equation}
with $l \in \{1,3,5\}$.
Specialising on K and H we obtain the representations
\begin{align}
\widehat{Q}_{6,p}^2  \psi_{l}(K) &= e^{-i\pi\frac{l}{3}}  \psi_{l}(K), \nonumber\\
  \widehat{Q}_{6,p}^2  \psi_{l}(H) &= e^{-i \pi \frac{l-p}{3}} \psi_{l}(H).
  \end{align}
In order to lift these representations to corepresentations of the isotropy groups of K and H we need to notice that the antiunitary operator 
$Q^3_{6,p}\mathbb{T}$ has a special feature in this 1-cell. Since $Q^3_{6,p}$ and $\mathbb{T}$ commute we know that $(Q^3_{6,p}\mathbb{T})^2=e^{-k_zp}$ and therefore the antiunitary operator $Q^3_{6,p}\mathbb{T}$
 squares to 1 on K for any $p$ and on H whenever $p$ is even, and squares to -1 on H  whenever $p$ is odd; see equation \eqref{Q3T on kz pi}.

On K the pair of representations $\{\psi_1(K),\psi_5(K)\}$ define an irreducible corepresentation of the third type while $\psi_3(K)$ defines a corepresentation of the first type. Note that Kramer's degeneracy rule does not apply to the state
$\psi_3(K)$ since the antiunitary operator $Q^3_{6,p}\mathbb{T}$ squares to 1 on K.

On H the relations of the representations depend on $p$.
For $p=1$ we have that $Q^3_{6,p}\mathbb{T}$ squares to -1 and all the states come in pairs.
The pair $\{\psi_3(H),\psi_5(H) \}$ defines a corepresentation of the third type while $\psi_1(H) \oplus \psi_1(H)$ defines a corepresentation of the second type. The combinatorial structure appears in diagram \eqref{KPH P61}.

\begin{equation} \label{KPH P61}
\xymatrix@1@R=0.2cm@C=1.5cm{
K &P& H\\
{{} \atop e^{i\pi3/3}}  \bullet \ar@{-}[rr]^(.5){\psi_3}  &&\bullet {e^{i\pi2 /3} \atop e^{i\pi 4/3}}\\
{e^{i\pi 5/3} \atop e^{i\pi 1/3}} \bullet \ar@{-}[rr]^(.8){\psi_{1}} \ar@{-}[rru]^(.2){\psi_5} && \bullet {e^{0} \atop e^{0}}\\
{e^{i\pi 1/3} \atop e^{i\pi 5/3}} \bullet \ar@{-}[rr]^(.8){\psi_{5}} \ar@{-}[rru]^(.2){\psi_1} && \bullet {e^{i\pi 4/3} \atop e^{i\pi 2/3}}\\
{e^{i\pi 3/3} \atop {}} \bullet  \ar@{-}[rru]^(.2){\psi_3} && 
}
\end{equation}
The minimal set of compatibility relations from BCS is:
\begin{align*}
    K_1(1) & \to P_1(1) &   P_1(1)\oplus P_2(1) & \leftarrow H_1H_3(2) \\
    K_2K_3(2) & \to P_2(1) \oplus P_3(1) &  2P_3(1) & \leftarrow H_2H_2(2) 
\end{align*}
which build into diagram 
\eqref{KPH P61 Bandrep}. This agrees with diagram  \eqref{KPH P61}.

\begin{equation} \label{KPH P61 Bandrep}
\xymatrix@1@R=0.2cm@C=1.5cm{
K_1(1)  \bullet \ar@{-}[rr]^(.5){P_1(1)}  &&\bullet H_1H_3(2)\\
K_2K_3(2) \bullet \ar@{-}[rr]^(.8){P_2(1)} \ar@{-}[rru]^(.2){P_2(1)} && \bullet H_2H_2(2)\\
K_2K_3(2) \bullet \ar@{-}[rr]^(.8){P_2(1)} \ar@{-}[rru]^(.2){P_3(1)} && \bullet H_1H_3(2)\\
K_1(1) \bullet  \ar@{-}[rru]^(.2){P_1(1)} && 
}
\end{equation}

For $p=2$ the operator $Q^3_{6,p}\mathbb{T}$ squares to 1 on H and therefore we have that $\{\psi_1(H),\psi_5(H) \}$ defines a corepresentation of the third type and $\psi_3(H)$ defines a corepresentation of the first type. The combinatorial structure appears in diagram \eqref{KPH P62}:

\begin{equation} \label{KPH P62}
\xymatrix@1@R=0.2cm@C=1.5cm{
K &P& H\\
{{} \atop e^{i\pi3/3}}  \bullet \ar@{-}[rr]^(.5){\psi_3}  &&\bullet {e^{i\pi1 /3} \atop e^{i\pi 5/3}}\\
{e^{i\pi 1/3} \atop e^{i\pi 5/3}} \bullet \ar@{-}[rr]^(.8){\psi_{5}} \ar@{-}[rru]^(.2){\psi_1} && \bullet {e^{i\pi 3/3} \atop {}}
}
\end{equation}

The minimal set of compatibility relations from BCS is:
\begin{align*}
    K_1(1) & \to P_1(1) &   P_1(1)\oplus P_3(1) & \leftarrow H_2H_3(2) \\
    K_2K_3(2) & \to P_2(1) \oplus P_3(1) &  P_1(1) & \leftarrow H_1(2) 
\end{align*}
which build into diagram 
\eqref{KPH P62 Bandrep}. This agrees with diagram  \eqref{KPH P62}.
\begin{equation} \label{KPH P62 Bandrep}
\xymatrix@1@R=0.2cm@C=1.5cm{
K_1(1)  \bullet \ar@{-}[rr]^(.5){P_1(1)}  &&\bullet H_2H_3(2)\\
K_2K_3(2) \bullet \ar@{-}[rr]^(.8){P_2(1)} \ar@{-}[rru]^(.2){P_3(1)} && \bullet H_1(1)
}
\end{equation}

For $p=3$ the operator $Q^3_{6,p}\mathbb{T}$ squares to -1 and therefore the pair $\{\psi_1(H), \psi_5(H) \}$ defines a corepresentation of the third type while $\psi_3(H) \oplus \psi_3(H)$ defines a corepresentation of the second type. The combinatorial structure appears in diagram \eqref{KPH P63}.

\begin{equation} \label{KPH P63}
\xymatrix@1@R=0.2cm@C=1.5cm{
K &P& H\\
{e^{i\pi 1/3} \atop e^{i\pi 5/3}} \bullet \ar@{-}@/^1pc/[rr]_(.7){\psi_1} \ar@{-}@/_1pc/[rr]^(0.3){\psi_5} &&\bullet {e^{i\pi 4/3} \atop e^{i\pi 2/3}}\\
{{} \atop e^{i\pi3/3}}  \bullet \ar@{-}[rrd]^(.5){\psi_3}  &&\\
{e^{i\pi 3/3} \atop {}} \bullet \ar@{-}[rr]^(.3){\psi_{3}} && \bullet {e^{0} \atop e^0}
}
\end{equation}

\begin{figure}
\includegraphics[width=8.5cm]{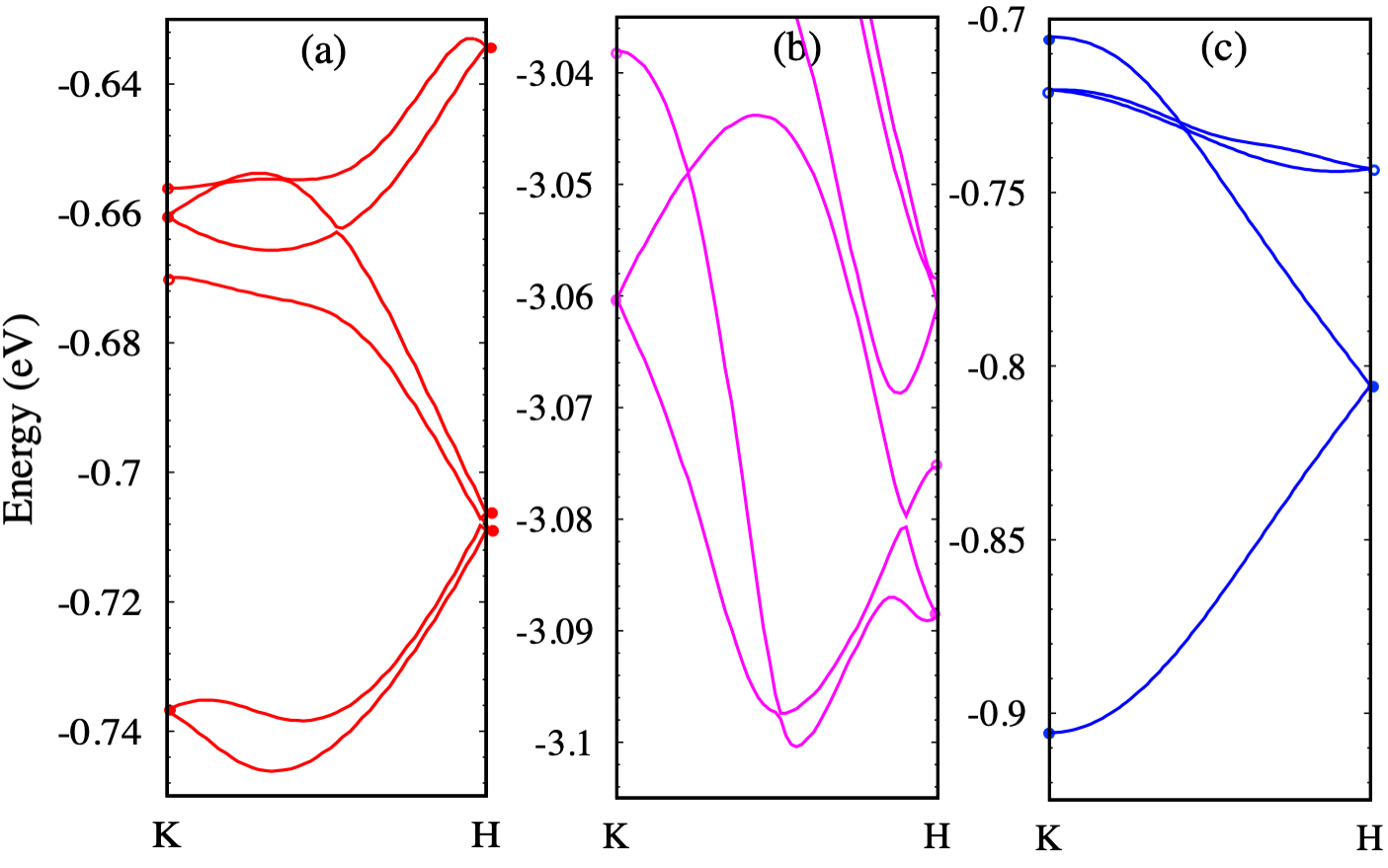}
\caption{Electronic band structure from K-H for a) P6$_1$ (In$_2$Se$_3$), b) P6$_4$ (KCaNd(PO$_4$)$_2$) and c) P$6_3$ (PI$_3$) space groups. The topological structure of the energy bands match the ones presented in diagrams \eqref{KPH P61}, \eqref{KPH P62} and \eqref{KPH P63} respectively.} \label{bands-KH}
\end{figure}

The minimal set of compatibility relations from BCS is:
\begin{align*}
    K_1(1) & \to P_1(1) &   P_2(1)\oplus P_3(1) & \leftarrow H_2H_3(2) \\
    K_2K_3(2) & \to P_2(1) \oplus P_3(1) &  2P_1(1) & \leftarrow H_1H_1(2) 
\end{align*}
which build into diagram  
\eqref{KPH P63 Bandrep} and agrees with diagram  \eqref{KPH P63}.
\begin{equation} \label{KPH P63 Bandrep}
\xymatrix@1@R=0.2cm@C=1.5cm{
K_2K_3(2) \bullet \ar@{-}@/^1pc/[rr]_(.7){P_2(1)} \ar@{-}@/_1pc/[rr]^(0.3){P_3(1)} &&\bullet H_2H_3(2)\\
K_1(1) \bullet \ar@{-}[rrd]^(.5){P_1(1)}  &&\\
K_1(1) \bullet \ar@{-}[rr]^(.3){P_1(1)} && \bullet H_1H_1(2)
}
\end{equation}
First principle calculations for hexagonal materials along the K-H path presented in Figure \ref{bands-KH} recover the combinatorial structure of the electronic energy bands.

\subsubsection{Topological Band analysis on A-L-H and $\Gamma$-M-K planes}

The isotropy group of all cells besides the 0-cells on the A-L-H and $\Gamma$-M-K planes is the group of order two $\langle \widehat{Q}^3\mathbb{T} \rangle$ generated by the
antiunitary operator $\widehat{Q}^3\mathbb{T}$.

Since $\widehat{Q}$ and $ \mathbb{T}$ commute we have that
\begin{equation}(\widehat{Q}^3\mathbb{T})^2 = \widehat{Q}^6 \mathbb{T}^2= e^{-ipk_z}\end{equation}
and therefore on the $\Gamma$-M-K plane and on the A-L-H plane whenever $p$ is even we have $(\widehat{Q}^3\mathbb{T})^2 =1$ and on the A-L-H plane 
 whenever $p$ is odd we have 
$(\widehat{Q}^3\mathbb{T})^2 =-1$.

The fact that $(\widehat{Q}^3\mathbb{T})^2 =1$ posses no restrictions on the possible representations of the energy bands along the plane $\Gamma$-M-K and on the plane A-L-H whenever $p$ is even. Nevertheless, whenever $p$ is odd, the fact that $(\widehat{Q}^3\mathbb{T})^2 =-1$ on the plane A-L-H implies by Kramer's degeneracy rule that all states come in pairs of equal energy. Therefore all energy bands along the paths A-L, L-H, and H-A are degenerate with degeneracy 2 whenever $p$ is odd.

This previous fact can be seen in the energy bands of the material PI$_3$ (P6$_3$, space group \#173) in Figure \ref{P63-bandas} where the energy bands along the paths A-L, L-H and H-A are degenerate, and on the paths $\Gamma$-M, M-K, and K-$\Gamma$ where there is no topological constraint.

\subsection{Symmetry groups $P6_p22$}

Whenever we compare the band representations of the symmetry groups $P6_p22$ and $P6_p$ along the high symmetry lines parallel to the six-fold rotation axis something quite interesting happens. It turns out that all minimal band representations along the high symmetry lines $\Gamma$-A, M-L, and K-H for the symmetry groups $P6_p22$ are isomorphic to the minimal band representations for the symmetry groups $P6_p$.

This interesting phenomenon is explained as follows. The irreducible corepresentations for the groups $P6_p22$ along $\Gamma$-A, M-L, and K-H restrict to the irreducible corepresentations for the groups $P6_p$ once the symmetry $M$ is forgotten. This restriction map is one-to-one and onto. Let us see with more detail this assertion.

The groups $P6_p22$ are obtained by adding to the groups P6$_p$ a $180^\circ$-degrees rotation along a specific axis; see  \eqref{list of generators in space coordinates}. In momentum space, these rotations behave as the operator
\begin{equation}M(k_x,k_y,k_z) = (k_y,k_x,-k_z)\end{equation}
and the commuting relation with $Q$ is $MQM=Q^{-1}$. Once lifted as operators 
on the Hilbert space of states we have the relations:
\begin{equation}\widehat{M}^2=-1, \ \ \widehat{M} \widehat{Q} \widehat{M}^{-1}= \widehat{Q}^{-1} \ \ \mbox{and} \ \ (\widehat{M}\widehat{Q})^2=-1,\end{equation}
which follow from the presence of SOC interaction.

\subsubsection{Topological band analysis on $\Gamma$-A}

On the $\Gamma$ and A points, if $Q$ has an eigenvector $\psi$ with eigenvalue $\lambda$ then $\widehat{M}$ is an eigenvector of $Q$ with eigenvalue $\lambda^{-1}$. Since  the order of $Q$ on $\Gamma$ and A is finite, then $|\lambda|=1$ and therefore $\lambda^{-1}= \overline{\lambda}$. Hence we have that 
the irreducible representations of $\langle \widehat{Q},\widehat{M} \rangle$ are of the form 
\begin{equation}
    \widehat{Q} \mapsto \left(
    \begin{array}{cc}
        \lambda & 0 \\
        0 & \lambda^{-1}
    \end{array}
    \right), \ \  \widehat{M} \mapsto \left(
    \begin{array}{cc}
        0 & 1 \\
        -1 & 0
    \end{array}
    \right)
\end{equation} 
where $\lambda$ is an appropriate root of unity. These representations
lift to corepresentations to the group $\langle \widehat{Q},\widehat{M}, \mathbb{T} \rangle$ by assigning to the operator $\widehat{M}\mathbb{T}$, which squares to 1,  the operator of complex conjugation $\mathbb{K}$. 
Note that this choice is coherent since
\begin{equation}(\widehat{M}\mathbb{T}) \widehat{Q} (\widehat{M}\mathbb{T})= \widehat{Q}^{-1} \ \ \mbox{and} \ \ (\widehat{M}\mathbb{T})^{-1} \widehat{M} (\widehat{M}\mathbb{T})= \widehat{M},\end{equation}
 and $\mathbb{K}$ flips the eigenvalues of $\widehat{Q}$ and commutes with the matrix associated to $\widehat{M}$.

Whenever $\lambda$ is not a real number,  the irreducible corepresentations
of the group $\langle \widehat{Q},\widehat{M}, \mathbb{T} \rangle$ are of the first kind, all of them are 2-dimensional and the eigenvalues of $\widehat{Q}$ on these corepresentations are the roots of unity $\lambda$ and $\overline{\lambda}$.
The restriction of these corepresentations to the group $\langle \widehat{Q}, \mathbb{T} \rangle$ become the corepresentations of the third type that were described in the analysis of the band diagram that appears in \eqref{GammaDeltaA P61}.

Whenever $\lambda$ is real the operator $\widehat{Q}$ commutes with $\widehat{M}$,  therefore the irreducible corepresentations are of the third type. Restricting these corepresentations to $\langle \widehat{Q}, \mathbb{T} \rangle$
we obtain the representations of the second type that were described in the analysis of the band diagram that appears in \eqref{GammaDeltaA P61}.

Now, on the $\Gamma$-A path the isotropy group is $\langle \widehat{Q}, \widehat{M} \mathbb{T} \rangle$.  Hence the 1-dimensional representations of $\widehat{Q}$ that appear in equation \eqref{eigenvectors Q GammaDeltaA line} lift to corepresentations of $\langle \widehat{Q}, \widehat{M} \mathbb{T} \rangle$ by assigning to $\widehat{M} \mathbb{T}$ the operator of complex conjugation $\mathbb{K}$. Therefore on the path $\Gamma$-A the description of the eigenvectors of $\widehat{Q}$ given in equation \eqref{eigenvectors Q GammaDeltaA line} works for the groups $P6_p22$ as well.

We conclude that the maximal band representations for the groups $P6_122$, $P6_222$ and $P_322$ along the path $\Gamma$-A  are the same as the ones the appear in figures \eqref{GammaDeltaA P61}, \eqref{GammaDeltaA P62} and \eqref{GammaDeltaA P63} respectively. 

\begin{figure}
\includegraphics[width=8.5cm]{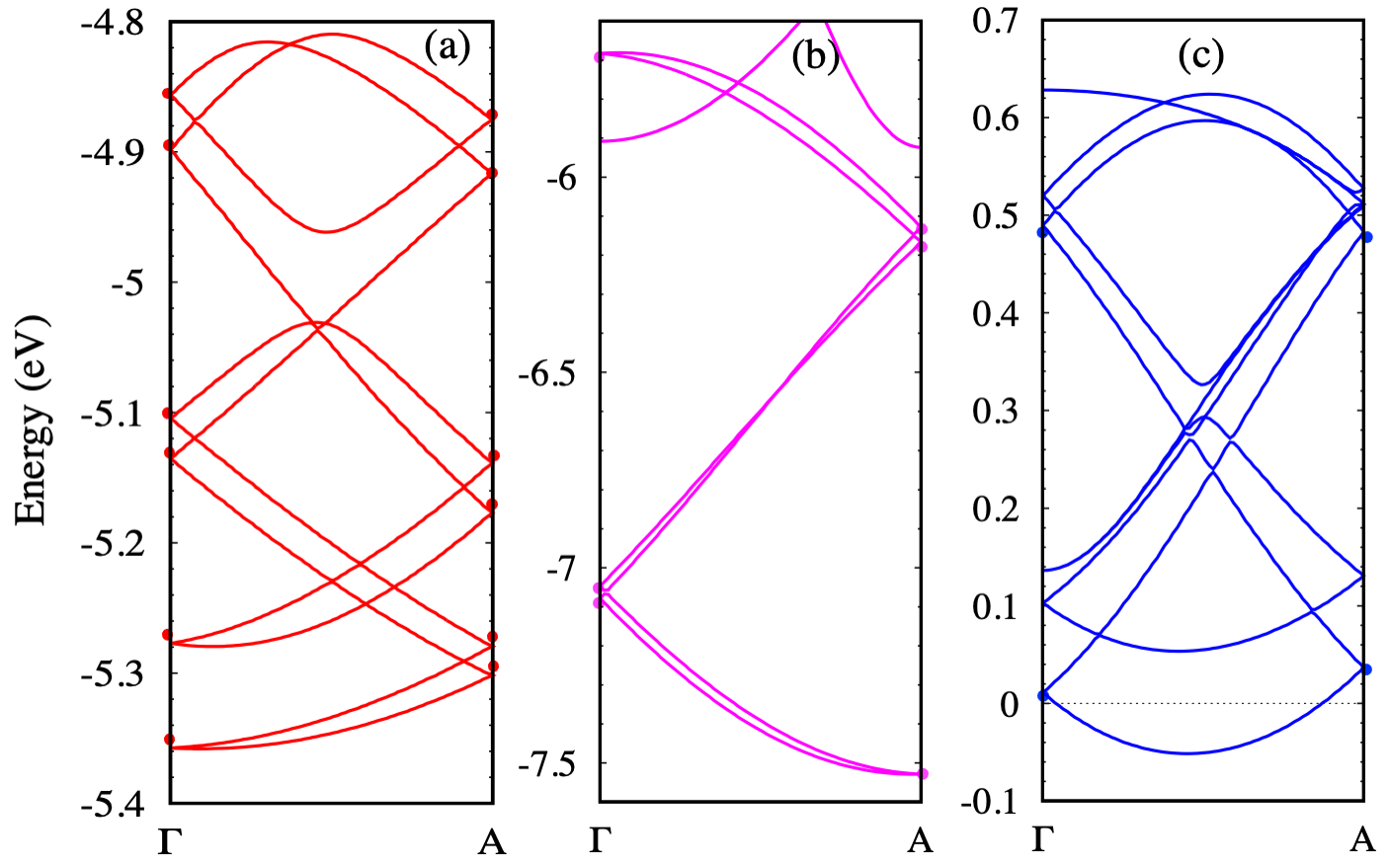}
\caption{Electronic band structure along $\Gamma$-A for a) P6$_1$22 (AgF$_3$), b) P6$_2$22 (TaGe$_2$) and c) P$6_3$22 (Nb$_3$CoS$_6$) space groups, The topological structure of the energy bands match with the ones presented in diagrams 
\eqref{GammaDeltaA P61}, \eqref{GammaDeltaA P62} and \eqref{GammaDeltaA P63} respectively.} \label{bands-GA222}
\end{figure}

First-principle calculations for hexagonal materials with $P6_p22$ space group symmetry presented in Figure \ref{bands-GA222} recover the combinatorial structure of the bands shown in \eqref{GammaDeltaA P61}, \eqref{GammaDeltaA P62} and \eqref{GammaDeltaA P63}. We can notice the formation of  Weyl points in the accordion-like (Figure \ref{bands-GA222}a) and hourglass-like dispersion (Figure \ref{bands-GA222}b and c), which are 
protected by 6-fold screw rotation symmetry.

\subsubsection{Topological band analysis on M-L}

The previous argument applies identically for the path M-L since we have that the isotropy group of M and L is $\langle \widehat{Q}^3, \widehat{M}\widehat{Q}, \mathbb{T} \rangle$ with $(\widehat{M}\widehat{Q})^2=-1$ and $(\widehat{M}\widehat{Q}) \widehat{Q}^3 (\widehat{M}\widehat{Q})^{-1}= \widehat{Q}^{-3}$. Here the operator $\widehat{M}\widehat{Q}$ plays the role that $\widehat{M}$ played in the $\Gamma$ and A points.

On the 1-cell M-L the isotropy group is  $\langle \widehat{Q}^3, \mathbb{T}\widehat{Q}^2\widehat{M}  \rangle$ and note that $(\mathbb{T}\widehat{Q}^2\widehat{M})^2=1$ and that $(\mathbb{T}\widehat{Q}^2\widehat{M}) \widehat{Q}^3 (\mathbb{T}\widehat{Q}^2\widehat{M}) = \widehat{Q}^{-3}$. Therefore the 1-dimensional representations of $\widehat{Q}^3$ defined for the group $P6_p$ on M-L in equation \eqref{eigenvectors Q MUL line} lift as
one-dimensional corepresentations of the group $\langle \widehat{Q}^3, \mathbb{T}\widehat{Q}^2\widehat{M}  \rangle$ whenever we represent the operator 
$\mathbb{T}\widehat{Q}^2\widehat{M}$ as the complex conjugation $\mathbb{K}$.

Hence the maximal band representations for the groups $P6_122$, $P6_222$ and $P_322$ along the 1-cell M-L are the same as the ones the appear in figures \eqref{MUL P61}, \eqref{MUL P62} and \eqref{MUL P63} respectively.

\begin{figure}
\includegraphics[width=8.5cm]{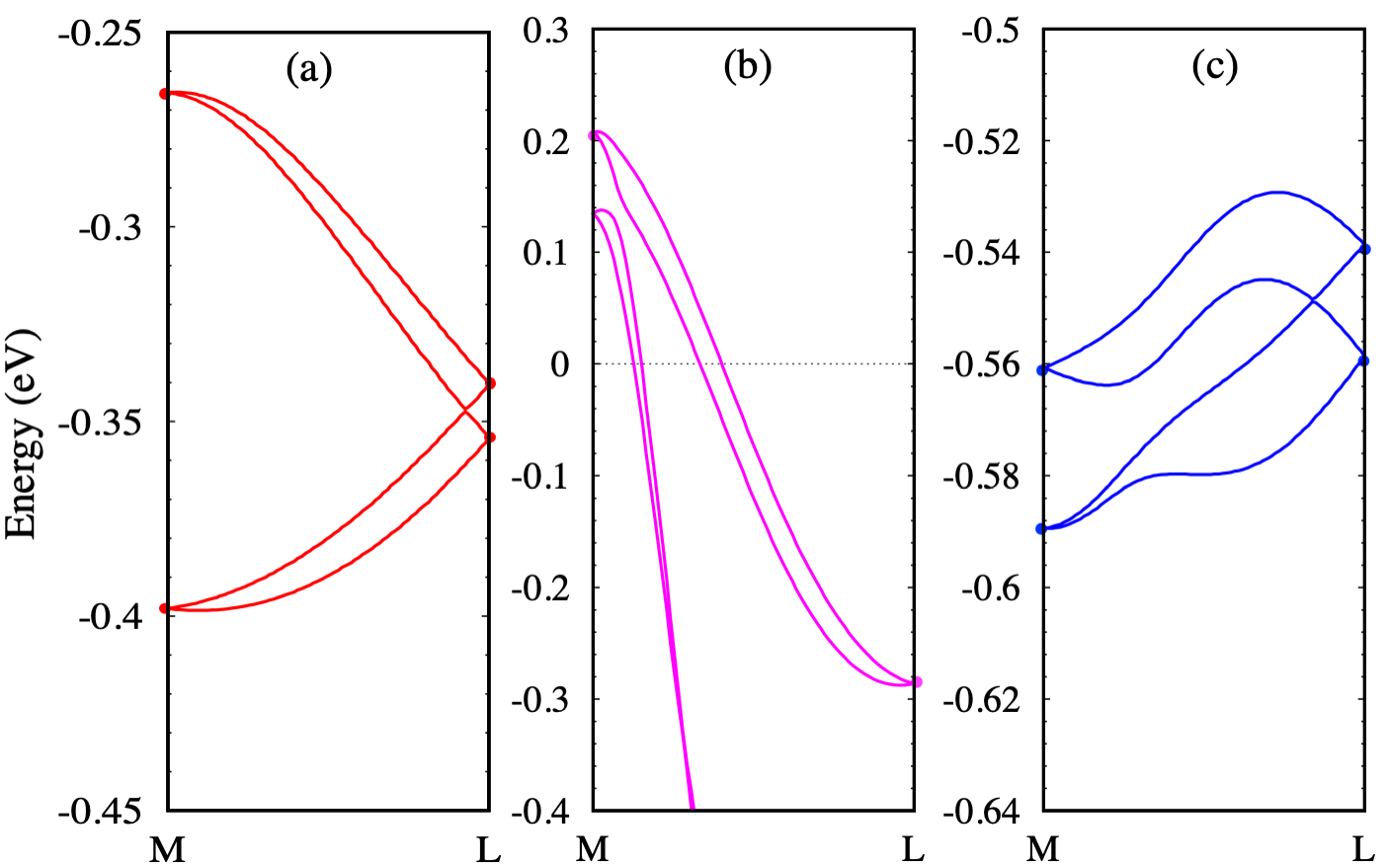}
\caption{Electronic band structure along M-L for a) P6$_1$22 (AgF$_3$), b) P6$_2$22 (TaGe$_2$) and c) P$6_3$22 (Nb$_3$CoS$_6$) space groups. The topological structure of the energy bands match the ones presented in diagrams \eqref{MUL P61}, \eqref{MUL P62} and \eqref{MUL P63} respectively.} \label{bands-ML222}
\end{figure}

Electronic band structure for $P6_p22$ materials along the M-L path presented in Figure \ref{bands-ML222} recover the combinatorial structures of the energy bands. From Figures \ref{bands-ML222}a and \ref{bands-ML222}c, we can see the formation of Weyl points with hourglass-like dispersion given by the relations presented in diagrams\eqref{MUL P61} and \eqref{MUL P63}.

\subsubsection{Topological band analysis on K-P-H}

The isotropy groups for K and H are generated by $\widehat{Q}^2$, $\widehat{M}$ and $\widehat{Q}^3\mathbb{T}$. On K we know that $(\widehat{Q}^3\mathbb{T})^2=1$ with  \begin{equation}(\widehat{Q}^3\mathbb{T}) \widehat{M} (\widehat{Q}^3\mathbb{T})= \widehat{M}^{-1}, \ \ (\widehat{Q}^3\mathbb{T}) \widehat{Q}^2 (\widehat{Q}^3\mathbb{T})= \widehat{Q}^2. \end{equation}
Whenever the eigenvalues of $\widehat{Q}^2$ in equation \eqref{eigenvectors Q KPH line} are not real, the 2-dimensional representation of $\langle \widehat{Q}^2, \widehat{Q}^3\mathbb{T} \rangle$ lifts to a representation of $\langle \widehat{Q}^2, \widehat{M}, \widehat{Q}^3\mathbb{T} \rangle$ as it was done before. 
Whenever the eigenvalue of $\widehat{Q}$ is real, the 1-dimensional representation of $\langle \widehat{Q}^2, \widehat{Q}^3\mathbb{T} \rangle$ may be lifted to one in $\langle \widehat{Q}^2, \widehat{M}, \widehat{Q}^3\mathbb{T} \rangle$ by assigning either $i$ or $-i$ as eigenvalue for $\widehat{M}$ and by representing the operator $\widehat{Q}^3\mathbb{T}$ by complex conjugation $\mathbb{K}$. Hence the 1-dimensional representation of $\langle \widehat{Q}^2, \widehat{Q}^3\mathbb{T} \rangle$ may be lifted to two different 1-dimensional representations of 
$\langle \widehat{Q}^2, \widehat{M}, \widehat{Q}^3\mathbb{T} \rangle$. 

On the point H whenever $p$ is odd the operator $\widehat{Q}^3 \mathbb{T}$ squares to -1. Therefore all 2-dimensional irreducible corepresentations of $\langle \widehat{Q}^2, \widehat{Q}^3\mathbb{T} \rangle$ lift to 2-dimensional irreducible corepresentations of the  group $\langle \widehat{Q}^2, \widehat{M}, \widehat{Q}^3\mathbb{T} \rangle$ as it was done above. Whenever $p$ is even the same argument as in the case of the point K can be carried out and the 1-dimensional corepresentations that appear lift to 1-dimensional representations of the group $\langle \widehat{Q}^2, \widehat{Q}^3\mathbb{T} \rangle$.

We conclude that the maximal band representations for the groups $P6_122$, $P6_222$ and $P_322$ along the 1-cell K-P-H are the same as the ones they appear in figures \eqref{KPH P61}, \eqref{KPH P62} and \eqref{KPH P63}, respectively. Here it is worth pointing out that the maximal band representation along the K-H line  for the groups P$6_122$ may appear in the shape of the following diagram:
\begin{equation}  \label{KPH P61 alternative}
\xymatrix@1@R=0.2cm@C=1.5cm{
K &P& H\\
 {e^{i\pi 5/3} \atop e^{i\pi 1/3}} \bullet \ar@{-}[rr]^(.5){\psi_5} \ar@{-}[rrd]^(.8){\psi_{1}}  &&\bullet {e^{i\pi4 /3} \atop e^{i\pi 2/3}}\\
{{e^{i\pi3/3}} \atop {}} \bullet  \ar@{-}[rru]^(.2){\psi_3} && \bullet {e^{0} \atop e^{0}}\\
{{} \atop {e^{i\pi 3/3}}} \bullet \ar@{-}[rr]^(.8){\psi_{3}}  && \bullet {e^{i\pi 2/3} \atop e^{i\pi 4/3}}\\
{e^{i\pi 1/3} \atop e^{i\pi 5/3}} \bullet  \ar@{-}[rru]_{\psi_5}  \ar@{-}[rruu]^(.2){\psi_1}&& 
}
\end{equation}
The open ends on the side of the K-point share the eigenvalue of -1 for the operator $\widehat{Q}^2$ and have respectively $+i$ and $-i$ as eigenvalues for $\widehat{M}$. These points will be joined  
when they reach the $\Gamma$-point, hence they will tend to appear close to each other.  

\begin{figure}
\includegraphics[width=8.5cm]{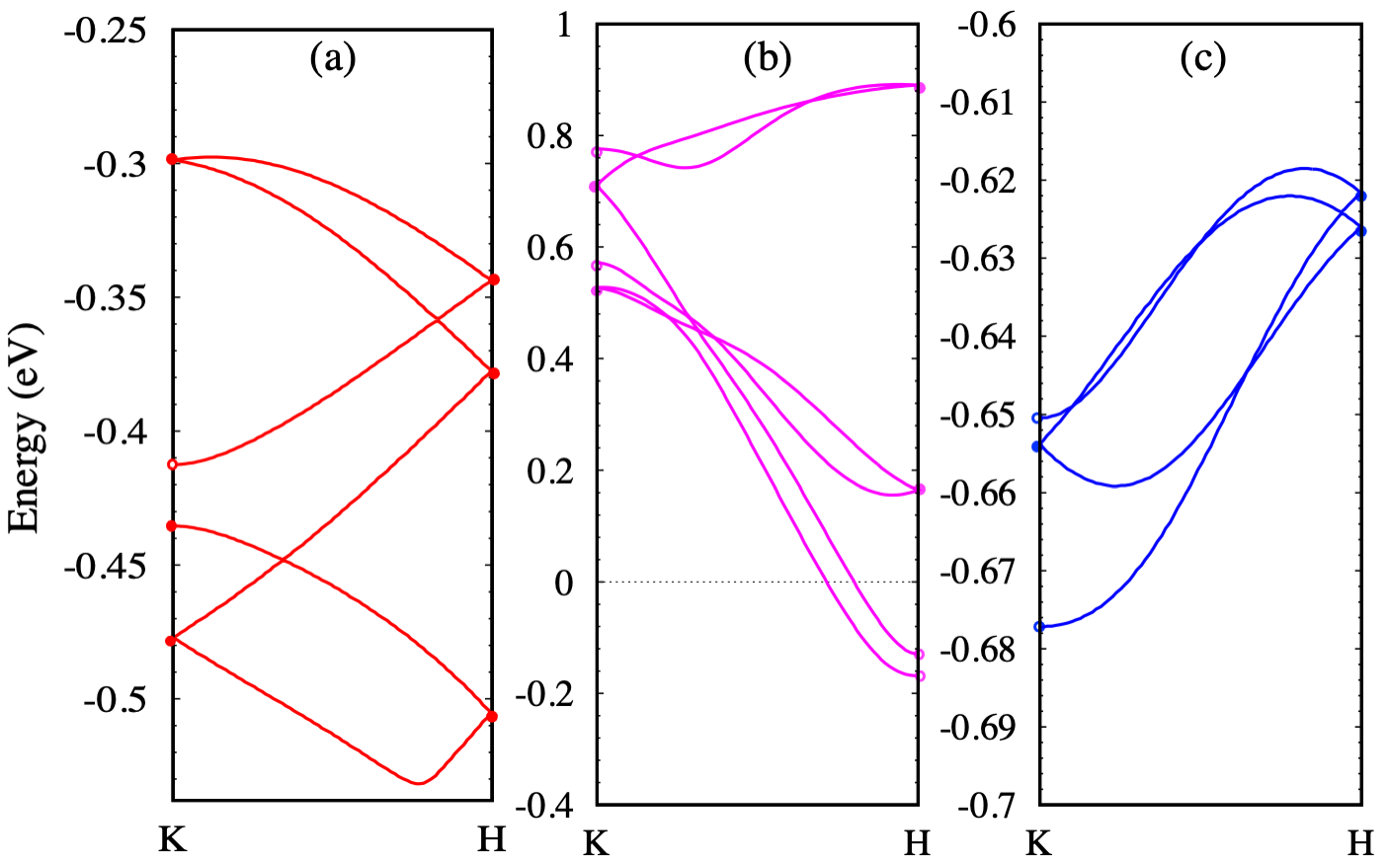}
\caption{Electronic band structure along K-H for a) P6$_1$22 (AgF$_3$), b) P6$_2$22 (TaGe$_2$) and c) P$6_3$22 (Nb$_3$CoS$_6$)  space groups. The topological structure of the energy bands match the ones presented in diagrams \eqref{KPH P61 alternative}, \eqref{KPH P62} and \eqref{KPH P63} respectively.}\label{bands-KH222}
\end{figure}

First-principle calculations for P6$_p$22 materials confirm the combinatorial structure of the energy bands along the K-H line as can be seen in Figure \ref{bands-KH222}. 

\subsubsection{Topological Band analysis on A-L-H and $\Gamma$-M-K planes}

The isotropy groups for the 1-cells on these planes consist of the groups $\langle \widehat{Q}^3 \mathbb{T}, \widehat{Q} \widehat{M} \rangle$, $\langle \widehat{Q}^3 \mathbb{T},  \widehat{M} \rangle$ and $\langle \widehat{Q}^3 \mathbb{T}, \widehat{Q}^3 \widehat{M} \rangle$. The unitary operators $\widehat{M}, \widehat{Q}\widehat{M}$ and $\widehat{Q}^3\widehat{M}$ square to -1 and their relation to the antiunitary operator $\widehat{Q}^3 \mathbb{T}$ is the following. On the $\Gamma$-M-K plane, and on the A-L-H plane whenever $p$ is even, we know that $(\widehat{Q}^3 \mathbb{T})^2=1$ and
\begin{equation}(\widehat{Q}^3 \mathbb{T}) \widehat{Q}^n \widehat{M} (\widehat{Q}^3 \mathbb{T})^{-1}=(\widehat{Q}^n \widehat{M})^{-1}.\end{equation}
Therefore the irreducible corepresentations are
of the first type and of dimension 1 in this case.

On the A-L-H plane whenever $p$ is odd the antiunitary operator obeys the equation $(\widehat{Q}^3 \mathbb{T})^2=-1$ and it commutes with the operators $\widehat{Q}^n \widehat{M}$, i.e.
\begin{equation}(\widehat{Q}^3 \mathbb{T}) \widehat{Q}^n \widehat{M} (\widehat{Q}^3 \mathbb{T})^{-1}=\widehat{Q}^n \widehat{M}.\end{equation}
Therefore the irreducible corepresentations are of the third type and of dimension 2, and all energy bands on the 1-cells A-L, L-H and A-H are degenerate with degeneracy of degree 2.

This could be observed on the energy bands of the materials AgF$_3$ and AuF$_3$ (P6$_1$22 \#178) in Figure \ref{P6122-bandas}  and CoNb$_3$S$_6$ (P6$_3$22 \#182) in Figure \ref{P6322-bandas} where the energy bands along the $\Gamma$-M, M-K and $\Gamma$-K paths are non-degenerate and along the paths A-L, L-H and A-H are degenerate of degree 2.

\subsection{Complete combinatorial band structure}

We may assemble the combinatorial diagrams described previously thus having a complete topological band structure. There are several ways to assemble the diagrams previously described and in what follows we will present some complete topological models that appear in the electronic band structure of the materials with the prescribed symmetry.

For the symmetry groups P$6_1$ and P$6_122$ we may assemble the topological band structure as it is shown in the following diagram:
\begin{equation} \label{complete topological band structure P6122}
\xymatrix@1@R=0.16cm@C=0.24cm{
\bullet \ar@{-}[r] & \bullet \ar@{-}[r] \ar@{-}[rdd] &  \bullet  \ar@{-}@/^0.4pc/[r] \ar@{-}@/_0.4pc/[r] &  \bullet \ar@{-}[r]  \ar@{-}[rdd] &  \bullet \ar@{-}[r] &  \bullet \ar@{-}[r]  \ar@{-}[rd] &  \bullet \ar@{-}@/^0.4pc/[r] \ar@{-}@/_0.4pc/[r]&  \bullet \ar@{-}@/^0.4pc/[r] \ar@{-}@/_0.4pc/[r]  &  \bullet \ar@{-}@/^0.4pc/[r] \ar@{-}@/_0.4pc/[r]&  \bullet \ar@{-}[r]  \ar@{-}[rdd] & \bullet \ar@{-}[r]  & \bullet \ar@{-}[rdd] \ar@{-}[r] & \bullet \\
 && && && \bullet  \ar@{-}[rd]&&&  \bullet  \ar@{-}[ru] && & \\
\bullet \ar@{-}[r] & \bullet \ar@{-}[r]  \ar@{-}[ruu] &  \bullet \ar@{-}@/^0.4pc/[r] \ar@{-}@/_0.4pc/[r]  &\bullet  \ar@{-}[rdd]  \ar@{-}[ruu]&  \bullet  \ar@{-}[r] &  \bullet \ar@{-}[ruu]  \ar@{-}[rdd]   &  &  \bullet \ar@{-}@/^0.4pc/[r] \ar@{-}@/_0.4pc/[r]&  \bullet \ar@{-}[ru] \ar@{-}[rd]&   & \bullet \ar@{-}[r] & \bullet \ar@{-}[ruu] \ar@{-}[r] & \bullet \\
&& && && \bullet  \ar@{-}[ru]&&& \bullet  \ar@{-}[rd] && & \\
\bullet \ar@{-}[r] & \bullet \ar@{-}[r] \ar@{-}[rdd]  &  \bullet  \ar@{-}@/^0.4pc/[r] \ar@{-}@/_0.4pc/[r] & \bullet  \ar@{-}[rdd]  \ar@{-}[ruu] & \bullet \ar@{-}[r] & \bullet \ar@{-}[r] \ar@{-}[ru]
 &  \bullet  \ar@{-}@/^0.4pc/[r] \ar@{-}@/_0.4pc/[r]  &  \bullet  \ar@{-}@/^0.4pc/[r] \ar@{-}@/_0.4pc/[r]  &  \bullet  \ar@{-}@/^0.4pc/[r] \ar@{-}@/_0.4pc/[r] & \bullet \ar@{-}[r] \ar@{-}[ruu]  & \bullet \ar@{-}[r] & \bullet \ar@{-}[rdd] \ar@{-}[r] & \bullet\\
 && && &&&&&&& & \\
\bullet  \ar@{-}[r] & \bullet \ar@{-}[r] \ar@{-}[ruu] &  \bullet  \ar@{-}@/^0.4pc/[r] \ar@{-}@/_0.4pc/[r] & \bullet  \ar@{-}[rdd] \ar@{-}[ruu] &  \bullet \ar@{-}[r] &  \bullet \ar@{-}[r]  \ar@{-}[rd] &  \bullet \ar@{-}@/^0.4pc/[r] \ar@{-}@/_0.4pc/[r]&  \bullet \ar@{-}@/^0.4pc/[r] \ar@{-}@/_0.4pc/[r]  &  \bullet \ar@{-}@/^0.4pc/[r] \ar@{-}@/_0.4pc/[r]&  \bullet \ar@{-}[r]  \ar@{-}[rdd] & \bullet \ar@{-}[r]  & \bullet \ar@{-}[ruu] \ar@{-}[r] & \bullet \\
 && && && \bullet  \ar@{-}[rd]&&&  \bullet  \ar@{-}[ru] && & \\
\bullet \ar@{-}[r] & \bullet \ar@{-}[r]  \ar@{-}[rdd]  &  \bullet \ar@{-}@/^0.4pc/[r] \ar@{-}@/_0.4pc/[r]  &\bullet  \ar@{-}[rdd]  \ar@{-}[ruu]&  \bullet  \ar@{-}[r] &  \bullet \ar@{-}[ruu]  \ar@{-}[rdd]   &  &  \bullet \ar@{-}@/^0.4pc/[r] \ar@{-}@/_0.4pc/[r]&  \bullet \ar@{-}[ru] \ar@{-}[rd]&   & \bullet \ar@{-}[r] & \bullet \ar@{-}[rdd] \ar@{-}[r] & \bullet  \\
&& && && \bullet  \ar@{-}[ru]&&& \bullet  \ar@{-}[rd] && &\\
\bullet \ar@{-}[r] & \bullet \ar@{-}[r]   \ar@{-}[ruu] &  \bullet  \ar@{-}@/^0.4pc/[r] \ar@{-}@/_0.4pc/[r] &  \bullet \ar@{-}[r] \ar@{-}[ruu] & \bullet \ar@{-}[r] & \bullet \ar@{-}[r] \ar@{-}[ru]
 &  \bullet  \ar@{-}@/^0.4pc/[r] \ar@{-}@/_0.4pc/[r]  &  \bullet  \ar@{-}@/^0.4pc/[r] \ar@{-}@/_0.4pc/[r]  &  \bullet  \ar@{-}@/^0.4pc/[r] \ar@{-}@/_0.4pc/[r] & \bullet \ar@{-}[r] \ar@{-}[ruu]  & \bullet \ar@{-}[r] & \bullet \ar@{-}[r] \ar@{-}[ruu] & \bullet \\
\mathrm{A} & \mathrm{L} & \mathrm{M} & \Gamma & \mathrm{A}& \mathrm{H} & \mathrm{K} & \Gamma & \mathrm{M} & \mathrm{K} & \mathrm{H} &\mathrm{L} & \mathrm{M}
}
\end{equation}
This structure could be seen on the twelve bands above the Fermi level in the material AgF$_3$ and AuF$_3$ as it is shown in Figure \ref{P6122-bandas}.  It is important to notice that any complete combinatorial band diagram for materials with symmetry groups P6$_1$ or
P6$_1$22 will contain multiple of 12 of bands. This is due to the combinatorial structure of the energy bands along the $\Gamma$-A line
as it is described in diagram \eqref{GammaDeltaA P61}. This fact agrees with the minimal insulating filling \cite{PhysRevB.94.155108, Watanabe14551} presented in \cite[Table III]{PhysRevLett.117.096404}
for the symmetry groups  P6$_1$ and
P6$_1$22 which in both cases is 12. The complete combinatorial band structure presented in \eqref{complete topological band structure P6122} is therefore minimal.

\begin{figure}
\includegraphics[width=8.9cm]{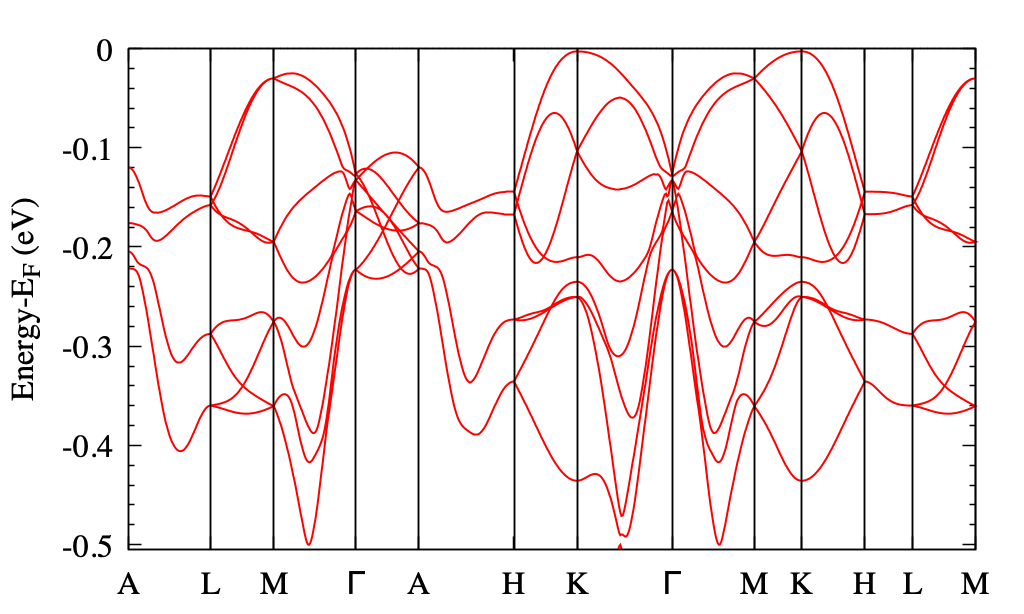}
\caption{Electronic band structure for PI$_3$ (P$6_3$ space group) with the combinatorial structure described in diagram \eqref{complete topological band structure P63}.
}
\label{P63-bandas}
\end{figure}

In the case of the symmetry group P$6_3$ the combinatorial structure of the energy bands can be assembled as described in the
following diagram:
\begin{equation} \label{complete topological band structure P63}
\xymatrix@1@R=0.16cm@C=0.24cm{
\bullet \ar@{-}[r] & \bullet \ar@{-}[r] \ar@{-}[rdd] &  \bullet  \ar@{-}@/^0.4pc/[r] \ar@{-}@/_0.4pc/[r] &  \bullet \ar@{-}[r]  \ar@{-}[rdddd] &  \bullet \ar@{-}[r] &  \bullet \ar@{-}[r]  \ar@{-}[rdd] &  \bullet \ar@{-}[r]&  \bullet \ar@{-}@/^0.4pc/[r] \ar@{-}@/_0.4pc/[r]  &  \bullet \ar@{-}[r] \ar@{-}[rd] &  \bullet \ar@{-}[r]   & \bullet \ar@{-}[r]  & \bullet \ar@{-}[rdd] \ar@{-}[r] & \bullet \\
 && && && \bullet  \ar@{-}[rd] \ar@{-}[ru]&&&  \bullet \ar@{-}@/^0.3pc/[rd] \ar@{-}@/_0.3pc/[rd] && & \\
\bullet \ar@{-}[r] & \bullet \ar@{-}[r]  \ar@{-}[ruu] &  \bullet \ar@{-}@/^0.4pc/[r] \ar@{-}@/_0.4pc/[r]  &\bullet  \ar@{-}[rdddd]  \ar@{-}[r]&  \bullet  \ar@{-}[r] &  \bullet  \ar@{-}@/^0.3pc/[ru] \ar@{-}@/_0.3pc/[ru] & \bullet \ar@{-}[r] &  \bullet \ar@{-}@/^0.4pc/[r] \ar@{-}@/_0.4pc/[r]&  \bullet \ar@{-}[ru] \ar@{-}[r] & \bullet   \ar@{-}[ruu] & \bullet \ar@{-}[r] & \bullet \ar@{-}[ruu] \ar@{-}[r] & \bullet \\
&& && &&&&& && & \\
\bullet \ar@{-}[r] & \bullet \ar@{-}[r] \ar@{-}[rdd]  &  \bullet  \ar@{-}@/^0.4pc/[r] \ar@{-}@/_0.4pc/[r] & \bullet   \ar@{-}[r] \ar@{-}[ruuuu] & \bullet \ar@{-}[r] & \bullet \ar@{-}@/^0.3pc/[rd] \ar@{-}@/_0.3pc/[rd]
 &  \bullet  \ar@{-}[r]  &  \bullet  \ar@{-}@/^0.4pc/[r] \ar@{-}@/_0.4pc/[r]  &  \bullet  \ar@{-}[r] \ar@{-}[dr] & \bullet  \ar@{-}[rdd]  & \bullet \ar@{-}[r] & \bullet \ar@{-}[rdd] \ar@{-}[r] & \bullet\\
 && && && \bullet  \ar@{-}[ru] \ar@{-}[rd]&&& \bullet   \ar@{-}@/^0.3pc/[ru] \ar@{-}@/_0.3pc/[ru] && & \\
\bullet  \ar@{-}[r] & \bullet \ar@{-}[r] \ar@{-}[ruu] &  \bullet  \ar@{-}@/^0.4pc/[r] \ar@{-}@/_0.4pc/[r] & \bullet  \ar@{-}[r]  \ar@{-}[ruuuu] &  \bullet \ar@{-}[r] &  \bullet \ar@{-}[r]  \ar@{-}[ruu] &  \bullet \ar@{-}[r] &  \bullet \ar@{-}@/^0.4pc/[r] \ar@{-}@/_0.4pc/[r]  &  \bullet \ar@{-}[r] \ar@{-}[ru]&  \bullet \ar@{-}[r]   & \bullet \ar@{-}[r]  & \bullet \ar@{-}[ruu] \ar@{-}[r] & \bullet\\
 \mathrm{A} & \mathrm{L} & \mathrm{M} & \Gamma & \mathrm{A}& \mathrm{H} & \mathrm{K} & \Gamma & \mathrm{M} & \mathrm{K} & \mathrm{H} &\mathrm{L} & \mathrm{M}
}
\end{equation}
This combinatorial structure could be seen on the eight bands below the Fermi level in the material PI$_3$ as it is shown in Figure \ref{P63-bandas}. At least two features make this combinatorial structure very interesting, the superposition of the hourglass energy bands along the $\Gamma$-A path and the way the energy bands are joined on the K point. These features cannot be deduced from the corepresentations of the isotropy groups nor the combinatorial structure along the edges of the Brillouin zone. It is through material or model energetics how the connectivity manifest  in the band structure could be determined. 

This is just one of the possible ways that the topological energy bands may be assembled and it is important to notice that it is not minimal. The minimal connectivity for P6$_3$ may look as diagram \eqref{minimal band structure P63} where the amount of bands agrees with the
minimal insulating filling presented in \cite[Table III]{PhysRevLett.117.096404}
for the symmetry group  P6$_3$ which is 4.

\begin{equation} \label{minimal band structure P63}
\xymatrix@1@R=0.16cm@C=0.24cm{
\bullet \ar@{-}[r] & \bullet \ar@{-}[r] \ar@{-}[rdd] &  \bullet  \ar@{-}@/^0.4pc/[r] \ar@{-}@/_0.4pc/[r] &  \bullet \ar@{-}[r]  \ar@{-}[rdd] &  \bullet \ar@{-}[r] &  \bullet \ar@{-}[r]  \ar@{-}[rdd] &  \bullet \ar@{-}[r]&  \bullet \ar@{-}@/^0.4pc/[r] \ar@{-}@/_0.4pc/[r]  &  \bullet \ar@{-}[r] \ar@{-}[rd] &  \bullet \ar@{-}[r]   & \bullet \ar@{-}[r]  & \bullet \ar@{-}[rdd] \ar@{-}[r] & \bullet \\
 && && && \bullet  \ar@{-}[rd] \ar@{-}[ru]&&&  \bullet \ar@{-}@/^0.3pc/[rd] \ar@{-}@/_0.3pc/[rd] && & \\
\bullet \ar@{-}[r] & \bullet \ar@{-}[r]  \ar@{-}[ruu] &  \bullet \ar@{-}@/^0.4pc/[r] \ar@{-}@/_0.4pc/[r]  &\bullet  \ar@{-}[ruu] \ar@{-}[r]&  \bullet  \ar@{-}[r] &  \bullet  \ar@{-}@/^0.3pc/[ru] \ar@{-}@/_0.3pc/[ru] & \bullet \ar@{-}[r] &  \bullet \ar@{-}@/^0.4pc/[r] \ar@{-}@/_0.4pc/[r]&  \bullet \ar@{-}[ru] \ar@{-}[r] & \bullet   \ar@{-}[ruu] & \bullet \ar@{-}[r] & \bullet \ar@{-}[ruu] \ar@{-}[r] & \bullet \\
 \mathrm{A} & \mathrm{L} & \mathrm{M} & \Gamma & \mathrm{A}& \mathrm{H} & \mathrm{K} & \Gamma & \mathrm{M} & \mathrm{K} & \mathrm{H} &\mathrm{L} & \mathrm{M}
}
\end{equation}

\begin{figure}
	\includegraphics[width=8.9cm]{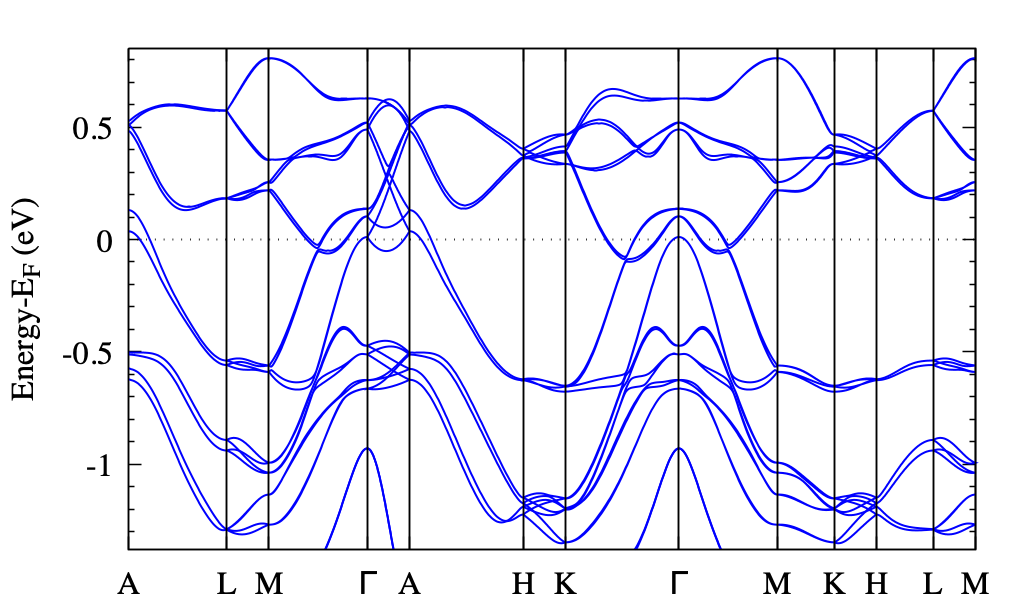}
	\caption{Electronic band structure for CoNb$_3$S$_6$ (P$6_3$22 space group) describing a complete
	combinatorial structure  with 20 energy bands.} \label{P6322-bandas}
\end{figure}

The topological assembly of the energy bands for the symmetry group P$6_322$ is more elaborate. It may require as much as 20 energy bands as it can be seen in the electronic band structure of CoNb$_3$S$_6$
that appears in Figure \ref{P6322-bandas}. The complete combinatorial diagram will not be included in this work.

\section{Appendix B\\ Material realisation}
\label{appendix:c}

We have used the materials project database \cite{materialsproject} and AFLOW database \cite{aflow} in order to show examples of materials for each of the space groups studied in the topological analysis. We have chosen materials with nonsymmorphic space groups generated by a $6$-fold rotation symmetry with $p$-screw symmetry operation. Moreover, we have compared these materials with the ones whose symmetry groups also include two extra symmetries of 2-fold rotation axis  around of axis which is perpendicular to the main $z$-axis (P6$_p$22 space groups).  

The materials used as examples for the prediction of the electronic band structure were: In$_2$Se$_3$, KCaNd(PO$_4$)$_2$, PI$_3$, AgF$_3$, AuF$_3$, TaGe$_2$ and Nb$_3$CoS$_6$. 

For P6$_1$ (\#169) we have used the compound indium selenide (In$_2$Se$_3$) in the $\gamma$-phase, which is called ``defect wurtzite structure" and it is a common crystal structure for In$_2$Se$_3$ thin-film epitaxial growth. In$_2$Se$_3$ is a direct semiconductor with a bandgap energy of around 1.0 eV and potential application in photovoltaic devices \cite{In2Se3}.

P$6_2$ (\#171) is not a common space group for the solid-state phase of normal materials. Instead we have used the space group P$6_4$ (\#172) which is band electronic topological equivalent to P6$_2$.  For P6$_4$ (\#172) we have used the double phosphate KCaNd(PO$_4$)$_2$ which is isotypic with the hexagonal phase of  LaPO$_4$. KCaNd(PO$_4$)$_2$ is an insulator material with a bandgap energy of $\sim$4.5eV which could be used in optoelectronic devices and white-light-emitting diodes when it is doped with transition-metal or rare-earth elements \cite{KCaNdPO4}.
 
For the P6$_3$ (\#173) space group we have used the phosphorus triiodide PI$_3$ which is an indirect semiconductor material ($\sim$2.0 eV) with valence band maximum at the $K$-point. The electronic band structure computed for PI$_3$ in Figure \ref{P63-bandas} shows the correspondence with the predicted group theory analysis presented in diagram \eqref{complete topological band structure P63}. We can notice the formation of 2, 2 and 3 Weyl points between the $k$-lines L-M, $\Gamma$-A, H-K  respectively. These Weyl points are symmetry protected by the 6-fold screw symmetry. We also notice a special energy crossing on the K-point which is characteristic of 6-fold symmetry systems like graphene and silicene. This crossing is predicted by the group theory analysis in diagram \eqref{complete topological band structure P63} as one of the possible ways the topological bands assemble.

The space group  P6$_1$22 (\#178) is a common space group for triflouride materials. Two of the materials are AgF$_3$ and AuF$_3$ which have an interesting band structure topology as it is shown in Figure \ref{P6122-bandas}. It is worth noting that in this case a complete topological band structure was determined in diagram \eqref{complete topological band structure P6122} from the band analysis carried out in the previous section. We can note an accordion-like dispersion in the $\Gamma$-$A$ path (see Figure \ref{bands-GA222}), which produces at least five crossings along this $k$-path four of them Weyl points. The evolution of these 12 bands along the high symmetry lines produces 3 and 4 symmetry protected Weyl points at the L-M and K-H paths respectively (see Figures \ref{bands-ML222} and  \ref{bands-KH222}).

 For the case of  P$6_2$22 (\#180) space group, we have studied the tantalum germanide (TaGe$_2$) material which is an intermetallic compound with 12 symmetry operations and a band structure with Weyl points as presented in Figures \ref{bands-GA222}, \ref{bands-ML222} and  \ref{bands-KH222}.  The electronic band structure in Figure  \ref{bands-GA222} shows hourglass-like Weyl points in the $\Gamma$-$A$ path which are protected by the nonsymmorphic screw symmetry.
 
 Finally, we have used as an example for the P$6_322$ (\#182) space group the material Nb$_3$CoS$_6$, which is a quasi-two-dimensional material with Co atoms inserted between Nb-S layers. It makes a sandwich structure in the sequence of Co-SNbS-Co and its unit cell is one of an hexagonal lattice. Moreover, the Co atoms are localised in a chiral position on the $z$-axis thus producing a nonsymmorphic screw symmetry ($C_2$ rotation plus $T_{\mathbf{a}/2}$ partial translation) for this compound.  The electronic band crossings for CoNb$_3$S$_6$ including SOC interaction are shown in Figure \ref{P6322-bandas} where we can see the formation of hourglass-like dispersion on the $\Gamma$-A and L-M paths at different energies.  In particular it is noted that the material is a metal with hourglass dispersion at the Fermi energy. These particular energy crossings are protected by the time-reversal and screw symmetry of the CoNb$_3$S$_6$ material.  We can also see the formation of nodal lines around the $\Gamma$-point on the $k_z$=0 plane, which is due to the band crossings at the Fermi level on the K-$\Gamma$-M path (see Figure \ref{P6322-bandas}).

For the six materials described above we have carried out first-principles calculations and we have reproduced the predicted crossing points on the energy bands according to each space group symmetry.

\subsection{Computational Method}

In order to study the electronic band structure for the P6$_p$ and P6$_3$22 hexagonal materials, we carried out first-principles calculations within the density-functional theory (DFT) framework. Exchange and correlation effects were treated with generalized gradient approximation (GGA) \cite{pbe} as implemented in the Vienna ab-initio simulation package (VASP) \cite{vasp}. The calculations of spin-orbit coupling (SOC) interaction were included self-consistently at the DFT level. Electron wave function was expanded in plane waves up to cut-off energy of 500 eV and a grid of 0.02 ($2\pi$/\AA)   $k$-space resolution has been used to sample the first Brillouin zone (FBZ). For the hexagonal materials, we have used FINDSYM code \cite{findsym} to determine the correct crystal symmetry operations. As it is shown in Figures \ref{bands-GA}, \ref{bands-ML}, \ref{bands-KH}, \ref{bands-GA222}, \ref{bands-ML222},  \ref{bands-KH222}, \ref{P6122-bandas}, \ref{P63-bandas} and \ref{P6322-bandas}, we have found full agreement between the different theoretical approaches to the band structure calculation. Moreover, we have studied physics beyond the topological band crossings using the Wannier representation generated by the Wannier interpolation technique \cite{wannier90}. We have used the Wanniertools package \cite{wanniertools} in order to calculate the position and chirality (by using the Wilson-loop method) of possible Weyl points for the P6$_1$22 space group. We also calculated the Berry curvature, the $k$-resolved  ($240^3$ $k$-mesh) anomalous Hall conductivity \cite{wanniertools} and the spin Hall conductivity \cite{SHEcode} in order to clarify the contribution of selected Weyl points in the magneto-transport properties of these hexagonal materials.

\bibliographystyle{naturemag}
\bibliography{topological}

\end{document}